\documentclass[10pt,journal,twoside,final]{IEEEtran}

\usepackage{diagbox}
\usepackage{rotating}
\usepackage{graphicx,subfig}
\usepackage{cite}
\usepackage{color}
\usepackage{bigstrut}
\usepackage{multirow}
\usepackage{amsmath}
\usepackage{url}
\usepackage{array}
\newcolumntype{L}[1]{>{\raggedright\let\newline\\\arraybackslash\hspace{0pt}}m{#1}}
\newcolumntype{C}[1]{>{\centering\let\newline\\\arraybackslash\hspace{0pt}}m{#1}}
\newcolumntype{R}[1]{>{\raggedleft\let\newline\\\arraybackslash\hspace{0pt}}m{#1}}
\usepackage{amssymb}
\DeclareMathOperator*{\argmax}{argmax}

\newcommand{\etal}{\textit{et al. }}

\begin{document}

\title{Towards Length-Versatile and Noise-Robust Radio Frequency Fingerprint Identification}

\author{Guanxiong~Shen,
	Junqing~Zhang,
	Alan~Marshall,~\IEEEmembership{Senior Member,~IEEE},
	Mikko~Valkama,~\IEEEmembership{Fellow,~IEEE},
	and Joseph~Cavallaro,~\IEEEmembership{Fellow,~IEEE}	
	
	\thanks{Manuscript received xxx; revised xxx; accepted xxx. Date of publication xxx; date of current version xxx. The work was in part supported by UK Royal Society Research Grants under grant ID RGS\slash R1\slash 191241 and National Key Research and Development Program of China under grant ID 2020YFE0200600. This paper was presented in part at the Asilomar Conference on Signals, Systems, and Computers 2021.
		The review of this paper was coordinated by xxx. 
		\textit{(Corresponding author: Junqing Zhang.)}}
	\thanks{G.~Shen, J.~Zhang and A.~Marshall are with the Department of Electrical Engineering and Electronics, University of Liverpool, Liverpool, L69 3GJ, United Kingdom. (email: Guanxiong.Shen@liverpool.ac.uk; junqing.zhang@liverpool.ac.uk; alan.marshall@liverpool.ac.uk)}
	\thanks{M.~Valkama is with the Department of Electrical Engineering, Tampere University, 33720 Tampere, Finland (email: mikko.valkama@tuni.fi)}
	\thanks{J.~Cavallaro is with the Department of Electrical and Computer Engineering, Rice University, Houston, USA. (email: cavallar@rice.edu)}
	\thanks{Color versions of one or more of the figures in this paper are available online at http://ieeexplore.ieee.org.}
	\thanks{Digital Object Identifier xxx}	
}


%

\maketitle

\begin{abstract}


Radio frequency fingerprint identification (RFFI) can classify wireless devices by analyzing the signal distortions caused by the intrinsic hardware impairments. State-of-the-art neural networks have been adopted for RFFI. However, many neural networks, e.g.,  multilayer perceptron (MLP) and convolutional neural network (CNN), require fixed-size input data. In addition, many IoT devices work in low signal-to-noise ratio (SNR) scenarios but the RFFI performance in such scenarios is rarely investigated. In this paper, we analyze the reason why MLP- and CNN-based RFFI systems are constrained by the input size. To overcome this, we propose four neural networks that can process signals of variable lengths, namely flatten-free CNN, long short-term memory (LSTM) network, gated recurrent unit (GRU) network and transformer. We adopt data augmentation during training which can significantly improve the model's robustness to noise. We compare two augmentation schemes, namely offline and online augmentation. The results show the online one performs better. During the inference, a multi-packet inference approach is further leveraged to improve the classification accuracy in low SNR scenarios. We take LoRa as a case study and evaluate the system by classifying 10 commercial-off-the-shelf LoRa devices in various SNR conditions. The online augmentation can boost the low-SNR classification accuracy by up to 50\% and the multi-packet inference approach can further increase the accuracy by over 20\%.

\end{abstract}
	

\section{Introduction}

\IEEEPARstart{R}{adio} frequency fingerprint identification (RFFI) is a promising technique for authenticating Internet of things (IoT) devices. 
The analog front-end of wireless devices consists of hardware components such as oscillator, mixer, and power amplifier, etc. These components are rich in hardware impairments, whose specification parameters slightly deviate from their nominal values.  
The impairments are unique and hence can be extracted as identifiers, working in a similar manner as biometric fingerprints~\cite{zhang2021radio}.

As RFFI can be considered as a multi-class classification problem,  deep learning techniques have been extensively exploited recently~\cite{zhang2021radio,liu2021bidirectional,he2016deep,al2021deeplora,shen2021infocom,shen2021jsac,peng2019deep,merchant2018deep,yu2019robust,xie2021generalizable,roy2019rfal,soltani2020more,al2020exposing,ozturk2020rf,qian2021specific,das2018deep}. RFFI uses a neural network to directly classify received signals and predict the device label. 
Conventional techniques manually design feature extraction algorithms. In contrast, deep learning-based approaches can automatically extract discriminative features after sufficient training. The performance of deep learning-based RFFI systems usually outperforms traditional approaches~\cite{robyns2017physical, das2018deep}.

Although deep learning-powered RFFI systems have shown excellent classification performance, they are constrained by the fixed input length and low SNR.
Previous studies use deep neural networks (DNNs), convolutional neural networks (CNNs), or recurrent neural networks (RNNs) as the intelligent engine~\cite{al2020exposing,soltani2020more,hanna2020open,zhang2021radio}. However, DNN and CNN are designed to handle a certain length of inputs. After the input length is determined, they are unable to process either longer or shorter signals, which conflicts with the fact that many wireless packets are variable in length. To overcome this, some studies use fixed-length packet preamble/segment as model inputs~\cite{ozturk2020rf,roy2019rfal,qian2021specific,shen2021jsac,hanna2020open}. However, this rules out the payload part and the preamble length may not be fixed in some wireless protocols. The fixed-size input problem of DNN and CNN has been indicated in previous works but the reason and analysis are not given~\cite{merchant2019enhanced,al2020exposing}. RNN models are able to process variable-length inputs by design. However, this important property has never been exploited in previous RFFI works. 
In addition, the transmission power of IoT end nodes should always be kept as minimum as possible to reduce power consumption, making  received signals susceptible to noise. It is challenging to correctly identify devices from low SNR signals.

In this paper, we take LoRa/LoRaWAN, a popular low power wide area network (LPWAN) technology, as a case study to investigate the above two challenges. 
Although the design methodology is applicable to any RFFI system that needs to tackle variable inputs and low SNR conditions, and is not limited to LoRa.
LoRa defines the physical layer modulation while LoRaWAN specifies higher layer protocols such as the medium access control (MAC) layer and network architecture.
LoRaWAN leverages the adaptive data rate (ADR) mechanism that enables end-nodes to adjust transmission configurations on the fly. This makes the LoRa preamble length variable.  LoRa transmission power is low, and the long-range communication results in serious attenuation. 


This paper designs RFFI protocols that can tackle input data with variable lengths and proposes solutions for low SNR scenarios.
We investigate four neural network architectures that are not constrained by the input size. 
We leverage the data augmentation to train noise-robust models, and compare the performance of different augmentation strategies. Finally, we propose a multi-packet inference approach, which can significantly improve the classification accuracy in low SNR scenarios.
Experimental evaluation is carried out using 10 commercial-off-the-shelf LoRa devices and a USRP N210 software defined radio (SDR) platform.
Our contributions are highlighted as follows:
\begin{itemize}
    
    \item We summarize and discuss the input size constraint of previously used neural networks, such as DNN and CNN. Four length-versatile neural networks are proposed to overcome the constraint, namely flatten-free CNN, long short-term memory (LSTM) network, gated recurrent unit (GRU) network and transformer. They are all capable of classifying LoRa devices from variable-length preambles. Transformer can achieve satisfactory performance with minimum complexity.
To the best knowledge of the authors, this is the first time to explore how to apply the state-of-the-art transformer model to RFFI.
We also compare their performance with the previously used slicing technique, which is the state-of-the-art solution for inputs with variable-length. Our solution is shown to perform better at low SNRs than the slicing technique.  
    \item We investigate the effect of data augmentation on enhancing RFFI noise robustness. We compare the performance of online, offline and no augmentation strategies. The models trained with online augmentation outperform the rest. Taking flatten-free CNN as an example, the model trained with online augmentation, offline augmentation, and no augmentation achieves accuracies of near 90\%, 80\%, and 10\%, respectively, at 15~dB. 
    \item We propose a lightweight multiple-packet inference method that can significantly improve the classification accuracy in low SNR scenarios. Specifically, the accuracy can be boosted from 20\% to over 50\% at 0~dB, and from 60\% to about 90\% at 10~dB.
\end{itemize}
In our previous work~\cite{shen2021asilomar}, we designed the transformer model to process variable-length signals. We also investigated the effect of data augmentation and multi-packet inference on improving the low-SNR RFFI performance. In this paper, we have significantly extended our contribution by studying three additional length-versatile neural networks, namely flatten-free CNN, LSTM and GRU. We also compare their performance with the slicing/splitting technique.

The rest of this paper is organized as follows. Section II presents the background and motivations. Section III introduces the RFFI system in detail and Section IV shows the architectures of the proposed length-versatile neural networks. Section V provides experimental settings, results and discussion. Section VI is the comparison with the slicing technique.
Section VII introduces related work and Section VIII finally concludes the paper.

\section{Background and Motivations}\label{sec:system_overview}
\subsection{LoRa/LoRaWAN Primer}
LoRa uses chirp spread spectrum (CSS) as the physical layer modulation technique. There are several basic up-chirps at the beginning of a LoRa packet, named preambles. A baseband LoRa preamble is given as
\begin{equation} 
	x'(t) = A e^{j(-\pi Bt + \pi \frac{B}{T} t^2)} \quad (0 \leq t \leq T), 
\end{equation}
where $A$, $B$ and $SF$ denote signal amplitude, bandwidth and spreading factor, respectively. $T$ is the duration of a LoRa symbol, given as
\begin{equation}\label{equ:preamble_time}
    T = \frac{2^{SF}}{B}, 
\end{equation}
The bandwidth increases from 125~kHz to 500 kHz whenever the SF ranges from 7 to 12. In our configuration, the entire preamble part $x(t)$ consists of eight repeating $x'(t)$. 

\begin{figure}[!t]
	\centering
	\includegraphics[width=3.4in]{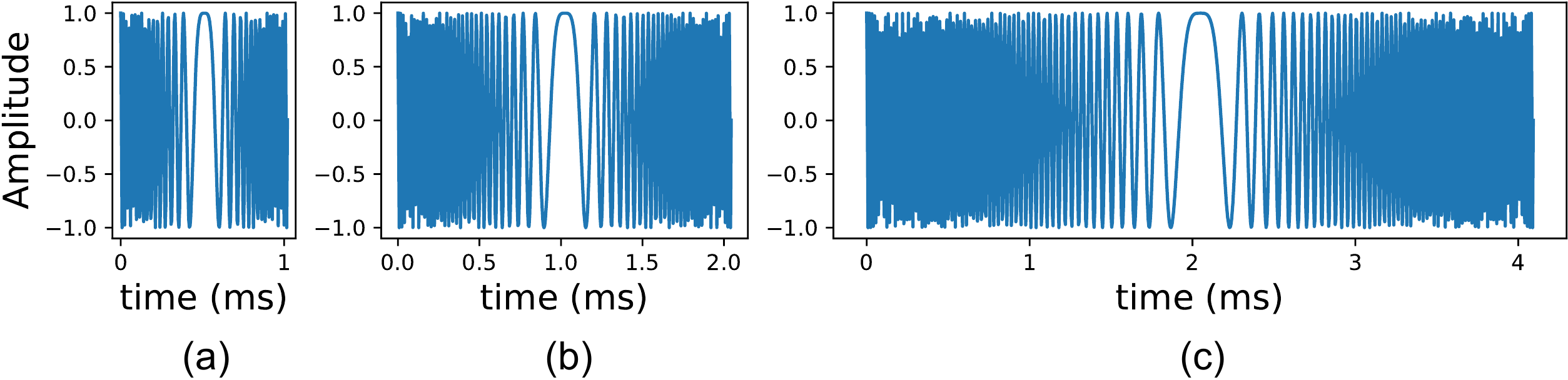}
	\caption{Waveform (in-phase branch) of a LoRa preamble under different SF configurations. The preamble length increases with SF. (a) SF = 7. (b) SF = 8. (c) SF = 9.}
	\label{fig:adr_waveform}
\end{figure}

The ADR mechanism is adopted in LoRaWAN, which allows LoRa end-nodes to optimize the data rate/SF adaptively according to the estimated SNR at the gateway. 
The SF will increase whenever a LoRa end-node moves further away from the gateway, in order to maintain a higher link budget~\cite{loraADR}.
As formulated in~(\ref{equ:preamble_time}), a higher SF leads to a longer LoRa symbol duration. Therefore, the LoRaWAN ADR mechanism makes the length of the preamble variable. Fig.~\ref{fig:adr_waveform} shows the waveform of one preamble under different SF configurations.

\subsection{RFFI}
\begin{figure}[!t]
	\centering
	\includegraphics[width=3.4in]{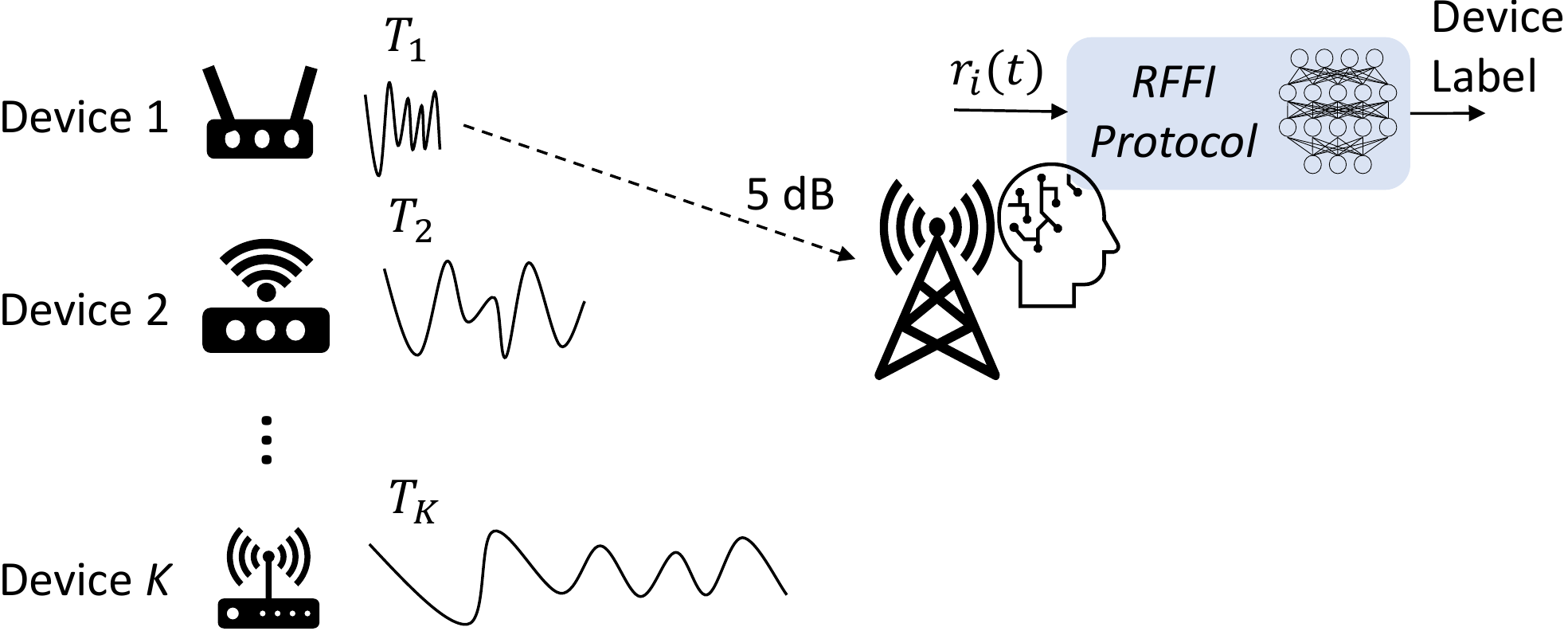}
	\caption{An RFFI system that can infer device labels from variable-length signals.}
	\label{fig:rffi_system}
\end{figure}

As shown in Fig.~\ref{fig:rffi_system}, an RFFI system involves $K$ IoT devices to be classified and a gateway as the authenticator. When the gateway receives a packet from a device, the RFFI protocol analyzes the received signal to infer from which end node the packet is sent.

For a specific IoT device~$i$, its signal $x(t)$ is distorted by hardware impairments and propagated in the wireless channel. At the receiver side, the received signal $r_i(t)$ is mathematically expressed as
\begin{align}
	r_i(t) = h(t)*\mathcal{F}_i(x(t))+v(t),
\end{align}
where $\mathcal{F}_i(\cdot)$ denotes the specific hardware distortion of device~$i$, $h(t)$ is the channel impulse response, $v(t)$ is the additive white Gaussian noise (AWGN) and $*$ denotes the convolution operation.
When the receiver captures a packet from device~$i$, it extracts the preamble part $r_i(t)$ and digitizes it, denoted as $r_i[n]$. Then an RFFI system is leveraged to accurately map this $r_i[n]$ to the device label $i$, acting as a classifier. This is formulated as
\begin{align}
	\mathcal{R}(r_i[n]) \longrightarrow i,
\end{align}
where $\mathcal{R}(\cdot)$ denotes operation of the RFFI system.

\subsection{Motivation 1: Variable-length Signal v.s. Fixed Input Length of Neural Networks}\label{sec:fixed_input_length}

Many previous deep learning-based RFFI systems can only process fixed-length inputs~\cite{al2020exposing,shen2021jsac,hanna2020open,robyns2017physical}. However, the length of $r_i(t)$ might be variable. For instance,  the lengths of LoRa preambles, i.e., $T_i$ in Fig.~\ref{fig:rffi_system}, are different due to the LoRaWAN ADR mechanism.





\subsubsection{Input Constraints of DNN}\label{sec:dnn_reason}
DNN, also known as fully connected neural network and multilayer perceptron (MLP), is used to build RFFI protocols in some previous works~\cite{robyns2017physical,roy2019rfal,shen2021infocom}, which only accept fixed-length inputs.

The DNN is entirely composed of dense layers. A dense layer defines a linear transformation between its input vector $\mathbf{x}$ and output vector $\mathbf{y}$, formulated as
\begin{equation}
    \mathbf{y} = \mathbf{x}\mathbf{A}^T + \mathbf{b}, 
\end{equation}
where $\mathbf{A}$ and $\mathbf{b}$ denote the weights and biases learned during training, respectively. The length of $\mathbf{x}$ must equal the number of rows of $\mathbf{A}^T$ to perform the multiplication. When an input longer than $\mathbf{x}$ is fed into the DNN, the matrix and vector dimension do not agree thus the first dense layer does not work. Therefore, DNN is constrained by the input size.

\subsubsection{Input Constraints of CNN}
\begin{figure}[!t]
	\centering
	\includegraphics[width=3in]{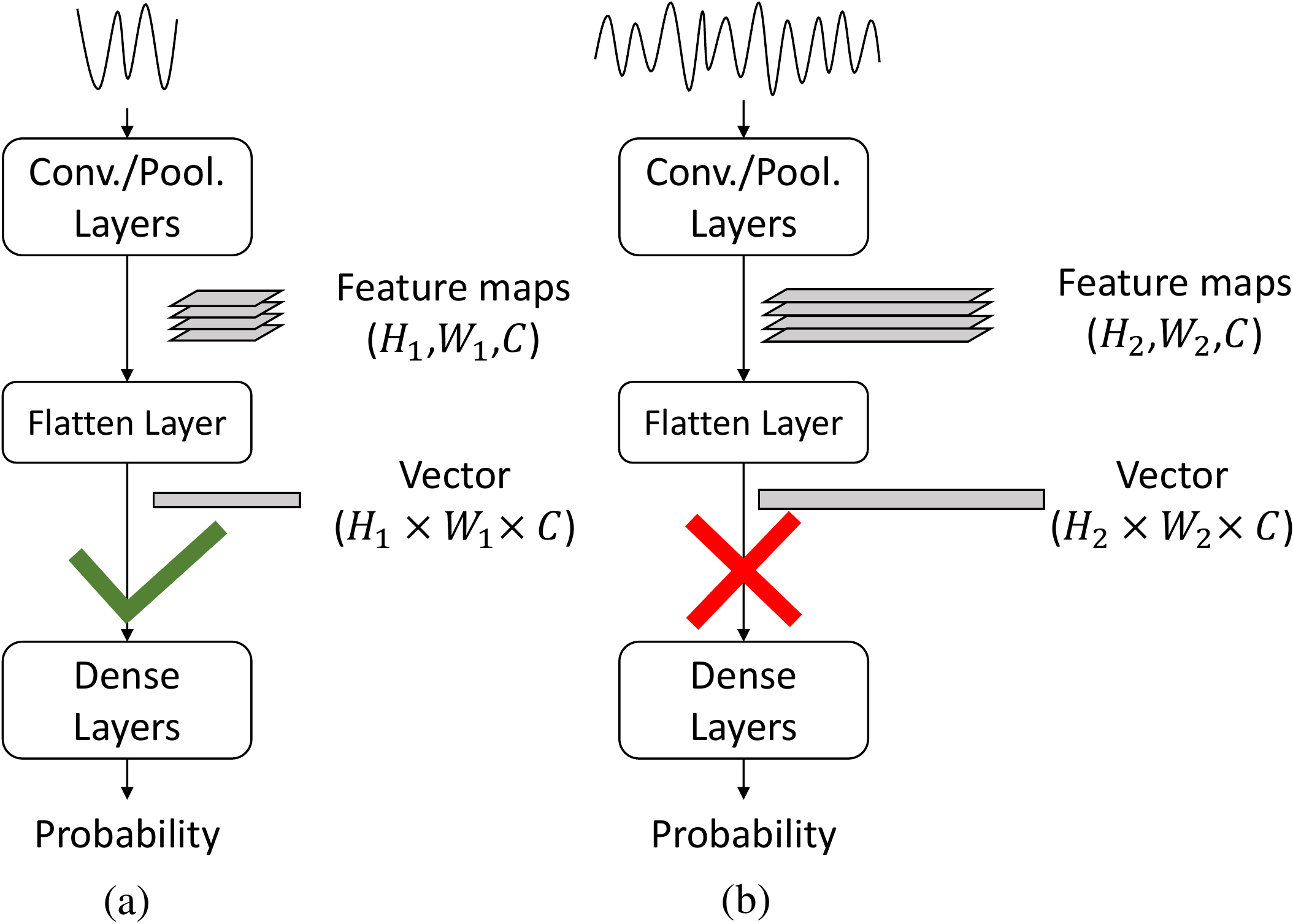}
	\caption{Fixed input size of CNN. (a) The CNN is trained with short signals. (b) The CNN cannot process longer signals as the flattened vector is longer than the signals during training. }
	\label{fig:cnn_reason}
\end{figure}
CNN is the most widely used model in RFFI. Previous studies have indicated that CNN can only handle signals of fixed size~\cite{al2020exposing,merchant2019enhanced,merchant2018deep}, but a proper explanation is missing. 

We show that CNN requires fixed size input due to the flatten and dense layers. CNN consists of two modules, convolutional/pooling layers for feature extraction and dense layers for classification. They are connected by a flatten layer which converts several two-dimension (2D) feature maps to a one-dimension (1D) vector. In fact, the feature extraction module is not sensitive to input length. The convolutional and pooling layers can process any size of the input, but will output feature maps of different sizes. This further makes the flatten layer output a vector of different lengths. However, as explained in Section~\ref{sec:dnn_reason}, dense layers are designed for a fixed input length, which constrains the input size of the entire CNN. 

As in the CNN exemplified in Fig.~\ref{fig:cnn_reason}(a), the input is first processed by several convolutional/pooling layers for feature extraction, and a $(H_1, W_1, C)$ tensor is returned. This denotes $C$ feature maps of size $(H_1, W_1)$, where $C$ equals the number of filters in the last convolutional layer. 
Since dense layers can only process 1D vectors, a flatten layer is employed to convert the $(H_1, W_1, C)$ tensor to a vector containing $(H_1 \times W_1 \times C)$ elements. The dense layers then make predictions from it.

However, the designed CNN cannot process signals of a different length. As shown in Fig.~\ref{fig:cnn_reason}(b), when a longer signal is input into the CNN, the output of convolutional/pooling layers is a $(H_2, W_2, C)$ tensor. Note that $H_2 \geq H_1$ and $W_2 > W_1$ since the input is longer. The flatten layer then converts the tensor to a  vector with $(H_2 \times W_2 \times C)$ elements, which cannot be processed by the dense layers. Therefore, the reason that CNN can only process fixed-size inputs is that dense layers cannot process variable-length vectors output from the flatten layer.


\subsection{Motivation 2: Low SNR of Received Signals}
RFFI systems should be able to identify wireless devices from low SNR signals. The SNR of $r_i[n]$ in the decibel scale is formulated as 
\begin{equation}
    \gamma  = 10\log_{10}\Big(\frac{RMS(h[n]* \mathcal{F}_i(x[n]))}{RMS(v[n])}\Big)
\end{equation}
where $RMS(\cdot)$ returns the root mean square value. The IoT devices are often configured with a low transmission power to save energy, resulting in a low amplitude emitted signal $\mathcal{F}_i(x[n])$. Moreover, a long distance between the transmitter and receiver results in significant attenuation, which affects the magnitude of channel taps in $h[n]$. These two factors jointly result in the low SNR of the received signal. It is necessary and challenging to improve RFFI performance in such low SNR scenarios.

\section{RFFI Protocol Design}

\subsection{Overview}

\begin{figure}[!t]
	\centering
	\includegraphics[width=3.4in]{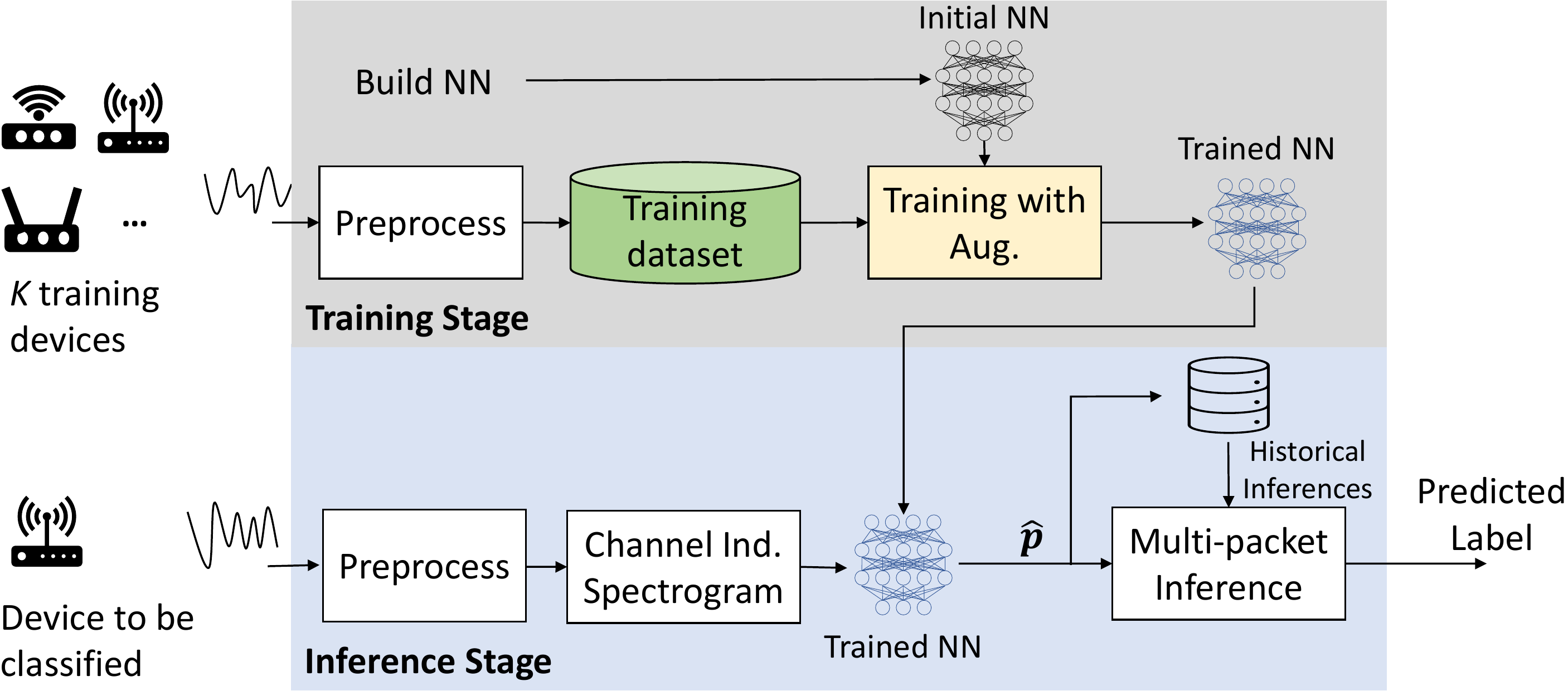}
	\caption{System overview.}
	\label{fig:system_overview}
\end{figure}

As shown in Fig.~\ref{fig:system_overview}, the proposed RFFI protocol involves two stages, namely training and inference. First, we collect extensive labelled LoRa packets from $K$ training devices, preprocess and store them as the training dataset. We also build a neural network that is capable of processing variable-size data and initialize its parameters. 
Then the neural network is trained with the collected dataset. Data augmentation is adopted during training to improve robustness to noise. Once the model training is completed, it can classify the newly received LoRa packet, i.e. the inference stage. The LoRa packet is preprocessed and converted to a channel independent spectrogram to mitigate channel effects. Then it is fed into the trained neural network and an inference is made.
In low SNR scenarios, we can leverage multiple packets for joint inference to obtain a more accurate prediction.


\subsection{Preprocessing}
As shown in Fig.~\ref{fig:system_overview}, the RFFI protocol needs to preprocess the received LoRa packet in both training and inference stages. We use the same preprocessing algorithms as in~\cite{shen2021towards}.

We employ the packet detection and synchronization algorithms to capture LoRa transmissions. After that, we extract the packet preamble part $r_i[n]$ for classification to prevent the model from learning protocol-specific information.
Then, carrier frequency offset (CFO) compensation is carried out as CFO is sensitive to temperature variation and reduces the system stability~\cite{shen2021infocom,shen2021jsac}.
Finally, the signal is normalized by dividing its root mean square, to prevent the model from classifying devices based on the received signal strength, which is location dependent. 





\subsection{Channel Independent Spectrogram}\label{sec:channel_ind_spectrogram}
During both training and inference stages, the received IQ samples $r_i[n]$ are converted to channel independent spectrograms to mitigate channel effects~\cite{shen2021towards}. Channel independent spectrogram is adopted because it has been demonstrated to be resilient to channel variations~\cite{shen2021towards}.

LoRa uses CSS modulation whose frequency changes over time. Hence short-time Fourier transform (STFT) is used to reveal how the frequency varies over time. A channel independent spectrogram, $\mathbf{S}^i$, can be further generated to mitigate channel effects, with each matrix element ${S}_{k,m}^{i}$ mathematically given as 
\begin{align}
		{S}_{k,m}^i &= 10\log_{10} \Big(\left | \frac{\sum_{n=0}^{N-1} r_i[n] w[n-mR]e^{-j2\pi \frac{k}{N} n}}{\sum_{n=0}^{N-1} r_i[n] w[n-(m-1)R]e^{-j2\pi \frac{k}{N} n}}  \right |^2 \Big)\nonumber\\ 
		&\mbox{for}\ k = 1,2,..., N\ \mbox{and}\ m = 1,..., M-1,
		\label{equ:channel_ind_spectrogram}
\end{align}
where $w[n]$ is a rectangular window of length $N$, and $R$ is the hop size. 
In this paper, $N$ and $R$ are always set to 64 and 32, respectively. Please refer to~\cite{shen2021towards} for the detailed derivation of the channel independent spectrogram.

\begin{figure}[!t]
	\centering
	\includegraphics[width=3.4in]{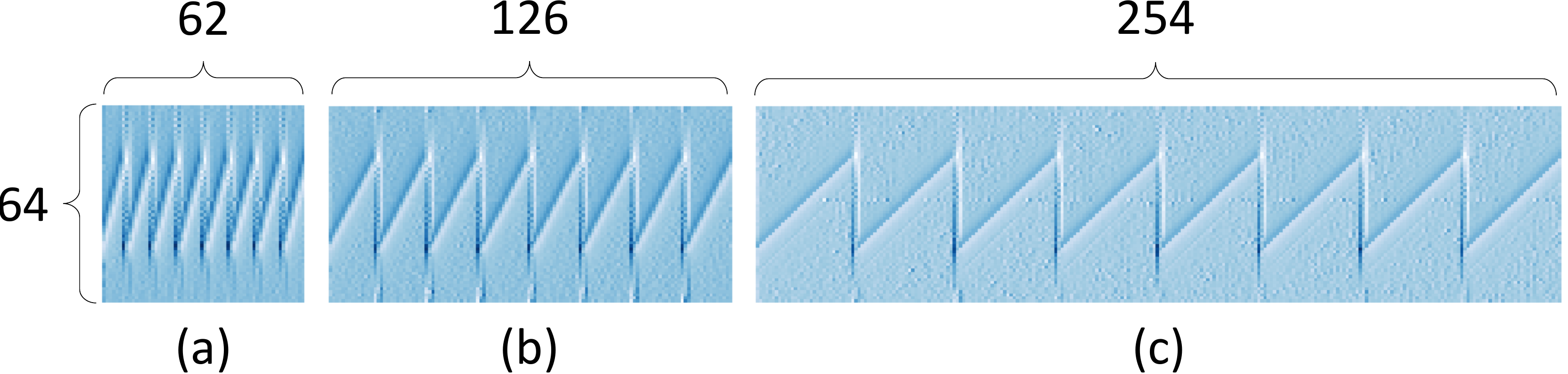}
	\caption{Channel independent spectrograms under different SF settings. (a) SF = 7. (b) SF = 8. (c) SF = 9.}
	\label{fig:channel_ind_spectrogram}
\end{figure}

The LoRa ADR mechanism introduced in Section~\ref{sec:system_overview} affects the size of channel independent spectrogram.
As formulated in (\ref{equ:channel_ind_spectrogram}), the generated channel independent spectrogram $\mathbf{S}^i$ is a $N\times (M-1)$ matrix. The height of the channel independent spectrogram, $N$, is equal to the length of the window function $w[n]$. It can be configured to be the same for all SFs. However, the width of $\mathbf{S}^i$, $(M-1)$, is related to the length of the received signal $r_i[n]$, which is formulated as
\begin{equation} 
	M-1 = \frac{8 \cdot \frac{2^{SF}}{B} \cdot \frac{1}{T_s}-N}{R},
\end{equation}
where $T_s$ is the receiver sampling interval. As shown in the formula, the width of $\mathbf{S}^i$ depends on the SF configuration $SF$, receiver sampling interval $T_s$, and transmission bandwidth $B$. 
Therefore, LoRaWAN ADR mechanism results in channel independent spectrograms $\mathbf{S}^i$ of different widths.
Fig.~\ref{fig:channel_ind_spectrogram} shows channel independent spectrograms when the SF is configured to 7, 8 and 9, respectively.
They have the same height but different widths. When SF is set to 9, the channel independent spectrogram is about four times as wide as when SF is set to 7. This demonstrates the input sizes of the neural network should not be fixed in LoRa-RFFI. Moreover, it validates the need for a neural network that can process channel independent spectrograms of variable size.

\subsection{Collecting Training Dataset and Building Neural Networks}
In the training stage, we collect a large number of LoRa packets, preprocess them and generate a training dataset. Note that we need to collect LoRa packets of various SF settings for training so that the system can handle different SFs in the inference stage. In this paper, we use SF 7, 8, and 9 as an example.

We also build a neural network by defining its architecture and initializing its parameters. The neural network architectures are described in detail in Section~\ref{sec:neural_network}. Note that all the neural networks can accept variable-length inputs therefore they can be trained with all the collected SF 7, 8 and 9 LoRa packets. After defining the architecture, the network parameters are initialized.

\subsection{Training with Augmentation}\label{sec:training_pipeline}

\begin{figure}[!t]
	\centering
	\includegraphics[width=3.4in]{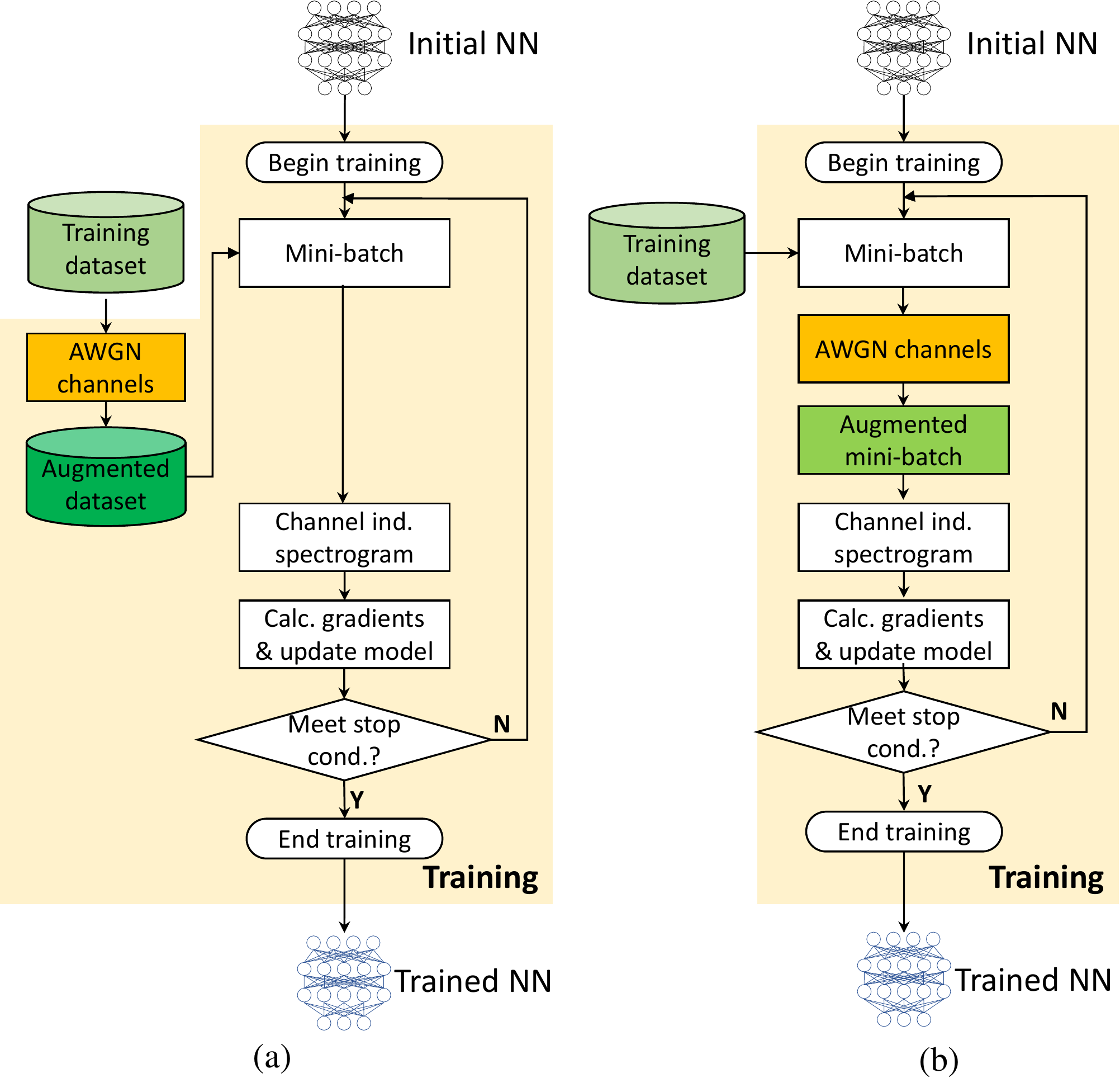}
	\caption{Training pipelines. (a) Offline augmentation. (b) Online augmentation.}
	\label{fig:training_pipelines}
\end{figure}

The parameters of the initial neural network can be trained and updated with the collected training packets.
We investigate two training pipelines with different augmentation strategies, namely offline and online augmentation, as shown in Fig.~\ref{fig:training_pipelines}.
Data augmentation is usually employed in the training stage of an RFFI protocol to increase the model robustness against channel variations~\cite{shen2021towards,merchant2019enhanced,al2021deeplora,soltani2020more}. In this paper, we focus on different SNR scenarios, therefore an AWGN channel model is employed for augmentation.

		


\subsubsection{Offline Augmentation}
The offline augmentation is performed on the original training dataset. We pass the training signals through AWGN channels with various SNR levels and store the augmented ones as the augmented training dataset. 
At the beginning of each epoch during training, a number of samples are randomly selected from the augmented training dataset to form a mini-batch. The IQ samples in the mini-batch are then converted into channel independent spectrograms and fed into the initial neural network. The gradients are calculated to update the neural network parameters. This update process continues until training stop conditions are met. In offline augmentation, the number of noisy signals learned during training equals the number of training samples. This can be increased by replicating the training dataset, but will be bounded by the storage space limit.

\subsubsection{Online Augmentation}
The online augmentation is also known as augmentation on-the-fly. Different from the offline one, online augmentation is performed on the mini-batches. As shown in Fig.~\ref{fig:training_pipelines}, the mini-batch is selected from the training dataset and fed into AWGN channels, obtaining an augmented mini-batch. The other training steps remain the same as the offline augmentation. In online augmentation, the model is trained with (steps $\times$ mini-batch size) noisy signals.

\subsection{Multi-packet Inference}
Once the training is completed, the neural network can act as a classifier to infer device identity. The received packet is preprocessed and converted to the channel independent spectrogram. Then it is fed into the trained neural network and the probability vector, $\mathbf{\hat{p}}$, will be returned by the softmax layer. $\mathbf{\hat{p}}$ can be regarded as confidence levels over $K$ devices. 

Multi-packet inference refers to making decisions with multiple LoRa transmissions. More specifically, we average the inference $\mathbf{\hat{p}}$ with $(N_{pkt}-1)$ historical inferences to derive a merged prediction $\mathbf{\hat{p}}^{'}$, mathematically expressed as
\begin{equation}\label{equ:multi_pkt}
    \mathbf{\hat{p}}{'} = \frac{1}{N_{pkt}}\sum_{n=1}^{N_{pkt}}  \mathbf{\hat{p}}^{n}.
\end{equation}
where $\mathbf{\hat{p}}^{n}$ is the prediction from the $n^{th}$ packet.
Then the predicted label can be derived by selecting the index with the highest probability, which is formulated as 
\begin{equation}
    label = \mathop{\argmax}_{k}(\mathbf{\hat{p}{'}}).
\end{equation}

According to the experimental results, the multi-packet inference protocol is particularly effective in low SNR scenarios. However, in high SNR scenarios, this protocol can be disabled to save computing resources. More detailed discussion can be found in Section~\ref{sec:experiment}.



\section{Length-Versatile Neural Networks}\label{sec:neural_network}
The neural network input is the channel independent spectrogram introduced in Section~\ref{sec:channel_ind_spectrogram}. As shown in Fig.~\ref{fig:channel_ind_spectrogram}, it has different widths at different SF settings. In this section, we will propose four models that can process channel independent spectrograms of any width, namely flatten-free CNN, LSTM, GRU and transformer.  

\subsection{Flatten-Free CNN}
\begin{figure*}[!t]
	\centering
	\includegraphics[width=6.8in]{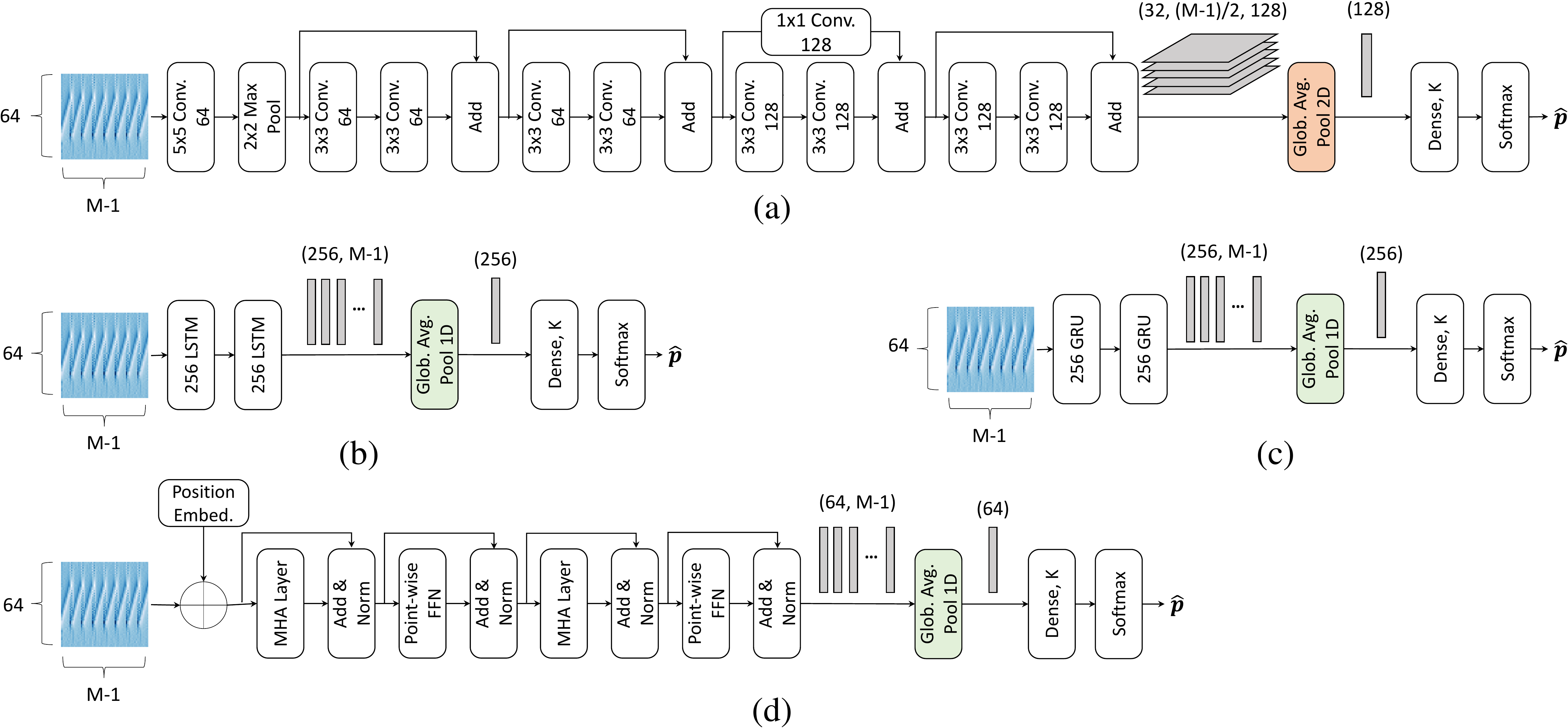}
	\caption{Architecture of the length-versatile neural networks. (a) Flatten-free CNN. (b) LSTM network. (c) GRU network. (d) Transformer.}
	\label{fig:networks}
\end{figure*}

As discussed in Section~\ref{sec:fixed_input_length}, CNN can only process fixed-size inputs because dense layers cannot process the flattened variable-length vectors. To overcome this, we replace the flatten layer with a global average pooling 2D layer so that the dense layers can always receive a fixed-length input. The architecture of the proposed flatten-free CNN is shown in Fig.~\ref{fig:networks}(a), which refers to the design of ResNet~\cite{he2016deep} but is more lightweight and optimized for the RFFI task. It consists of ten convolutional layers, one max pooling layer, one global average pooling 2D layer and one dense layer of $K$ neurons activated by softmax function. Skip connections are employed. 1x1 convolution is performed on the output of the fifth convolutional layer so that it can be added with the output with the seventh convolutional layer. The output of all the convolutional layers are zero-padded to be of the same dimension as the input, and then activated by ReLU function. 

The input channel independent spectrogram is fed into the convolutional and max pooling layers for feature extraction, and 128 feature maps of size (32, ($M-1)/2$) are returned, forming a (32, ($M-1)/2$, 128) tensor. The first two dimensions (32, ($M-1)/2$) depend on the size of input channel independent spectrogram and the third dimension equals the number of filters in the last convolutional layer.

A global average pooling 2D layer is leveraged to make the input to the dense layers independent of the channel independent spectrogram size. It is highlighted in orange in Fig.~\ref{fig:networks}(a). The global average pooling 2D layer calculates the average value of each individual feature map and outputs a fixed-length (128) vector that acts as the input to the dense layer. In other words, the size of dense layer input only depends on the number of filters in the previous convolutional layer. In this manner, regardless of the size of the input data, the dense layer can always receive input of fixed-length (128). 



\subsection{LSTM Network}

RNN is another popular class of deep learning models, which has been employed in some previous RFFI studies~\cite{shen2021jsac,al2021deeplora,das2018deep,roy2019rfal}.
RNN is designed for sequential data, such as speech and human language, whose length is naturally variable. Therefore, RNN models are not constrained by input size by design. However, this valuable property has never been discussed and investigated in previous RFFI studies. 

A popular variant of RNN is the LSTM network~\cite{hochreiter1997long}. RNN-structured neural networks can process variable-width channel independent spectrograms. The LSTM network is designed as shown in Fig.~\ref{fig:networks}(b), which consists of two 256-unit LSTM layers, one global average pooling 1D layer and one softmax-activated dense layer. The LSTM layers are configured to return the full sequence instead of the last output. Similar to convolutional layers, LSTM layers are insensitive to the input size as well. Therefore, how to make the input of the dense layer independent of the channel independent spectrogram width is the critical aspect of this design.

As shown in Fig.~\ref{fig:networks}(b), the channel independent spectrogram is processed by two LSTM layers and a tensor of shape (256, $M-1$) is obtained. The first dimension (features) equals the unit number 256 of the second LSTM layer, and the second dimension (timesteps) of the output tensor equals the width of the channel independent spectrogram, $M-1$. 
Therefore, the shape of the LSTM layers output (256, $M-1$) varies with the width of the input channel independent spectrogram. 

A global average pooling 1D layer is leveraged to make the dense layer input size constant, It is highlighted in green in Fig.~\ref{fig:networks}(b). It averages the tensor along the timesteps, and output a (256) vector. By this operation, the dense layers always receive a (256) vector regardless of the width of the input channel independent spectrogram.

\subsection{GRU Network}
GRU network is another variant of RNN, which can also handle variable-length inputs~\cite{chung2014empirical}. GRU units have less parameters compared to LSTM units. The designed GRU network is shown in Fig.~\ref{fig:networks}(c), whose architecture is nearly the same as the LSTM network, except that the LSTM layers are replaced by the GRU layers. Similar to LSTM layers, the GRU layers also output a (256, $M-1$) tensor and the global average pooling 1D layer can convert it to a fixed-length (256) vector, regardless of the input size. Therefore, GRU networks can handle variable-size inputs as well.

\subsection{Transformer}
In addition to RNN models, the transformer is another state-of-the-art model for processing variable-length sequential data~\cite{vaswani2017attention}.
As shown in Fig.~\ref{fig:networks}(d), the proposed transformer for RFFI consists of two multi-head attention (MHA) layers, two point-wise feed forward (point-wise FFN) layers, one global average pooling 1D layer and a dense layer with softmax activation. Similar to convolutional and LSTM layers, MHA layer and the point-wise FFN layer can also process input of any length. Hence the goal of transformer design is to make the input of dense layer a constant length.

The model design refers to the encoder part proposed in~\cite{vaswani2017attention}. First, the position embedding is added to the input channel independent spectrogram to enable the transformer to learn the temporal correlations. The result of addition is then fed into MHA and point-wise FFN layers for feature extraction. Skip-connections and layer normalization are leveraged to improve the feature learning ability. The output of the second point-wise FFN layer is a tensor of shape ($M-1$, 64), which is exactly the same size as the input channel independent spectrogram.
As shown in Fig.~\ref{fig:channel_ind_spectrogram}, channel independent spectrograms under different SF configurations have the same height $N$ = 64. Therefore, the tensor shape is simplified to (64, $M-1$), which is affected by the input width. 

Similar to the LSTM model, a global average pooling 1D layer is used to average the tensor along timesteps. It converts the (64, $M-1$) tensor to a (64) vector that acts as the input of dense layer. The vector is fixed-length regardless of the model input size, so the transformer can handle channel independent spectrograms with any SF configuration.

\subsection{Summary}
The critical design of a length-versatile classification neural network is to make the input to the dense layer constant, which is achieved by using global average pooling 2D/1D layers in our schemes. The global average pooling 2D layer is used in CNN models. It is placed after the convolutional layers, and converts the 3D tensor to a 1D vector by averaging each individual feature map. The output size of global average pooling 2D layer only depends on the number of filters in the last convolutional layer. The global average pooling 1D layer is used in RNN and transformer models. It averages along the time steps to convert the LSTM/GRU/FFN layer outputs to a constant-length vector. With the help of global average pooling layers, the dense layer always receives a fixed-length vector regardless of the model input size, thus is able to process variable-length received signals $r_i(t)$.


\section{Experimental Evaluation}\label{sec:experiment}
\subsection{Experimental Settings}
\subsubsection{Data Collection Settings}
\begin{figure}[!t]
	\centering
	\subfloat[]{\includegraphics[width=1.7in]{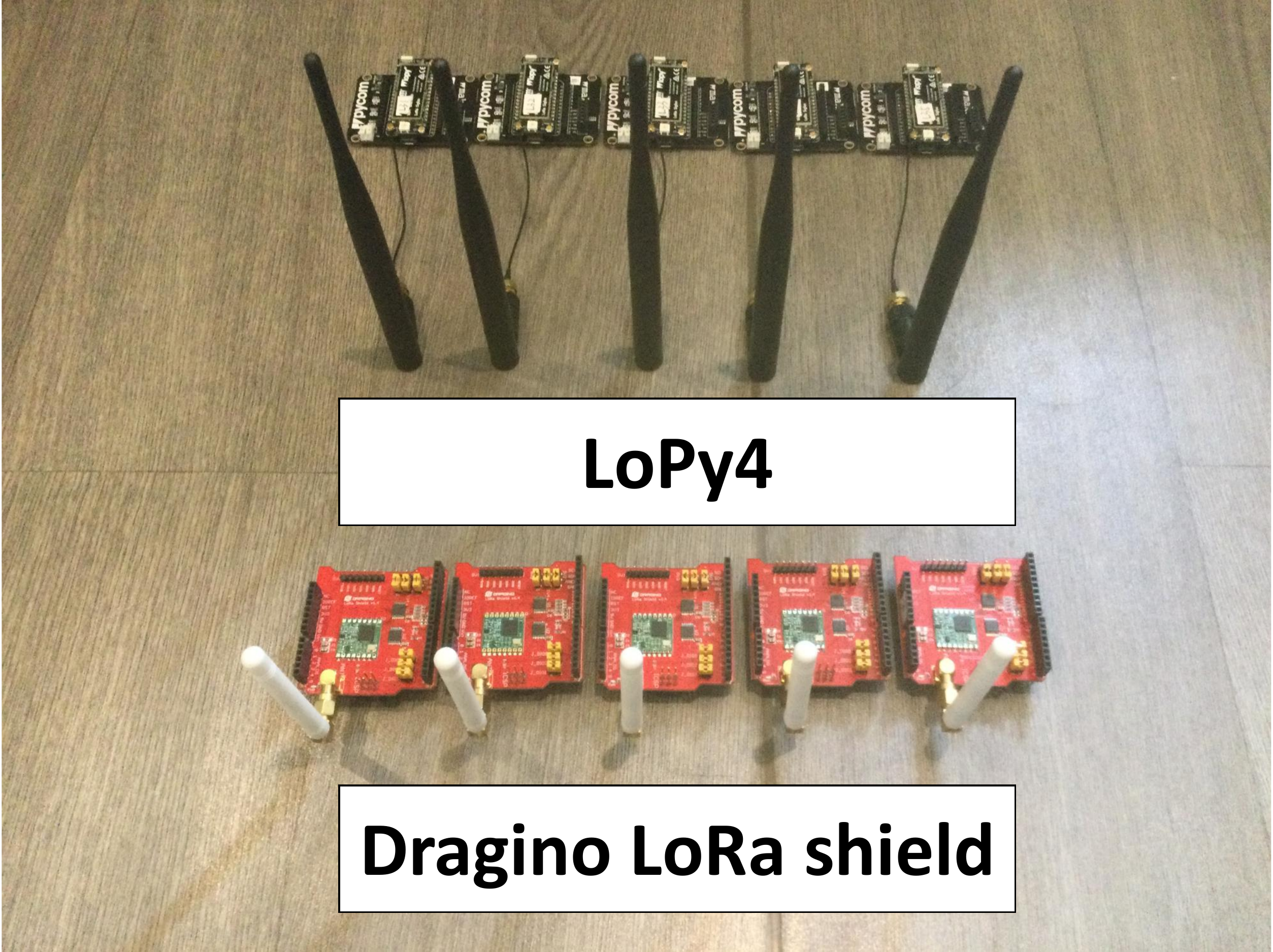}
		\label{}}
	\subfloat[]{\includegraphics[width=1.7in]{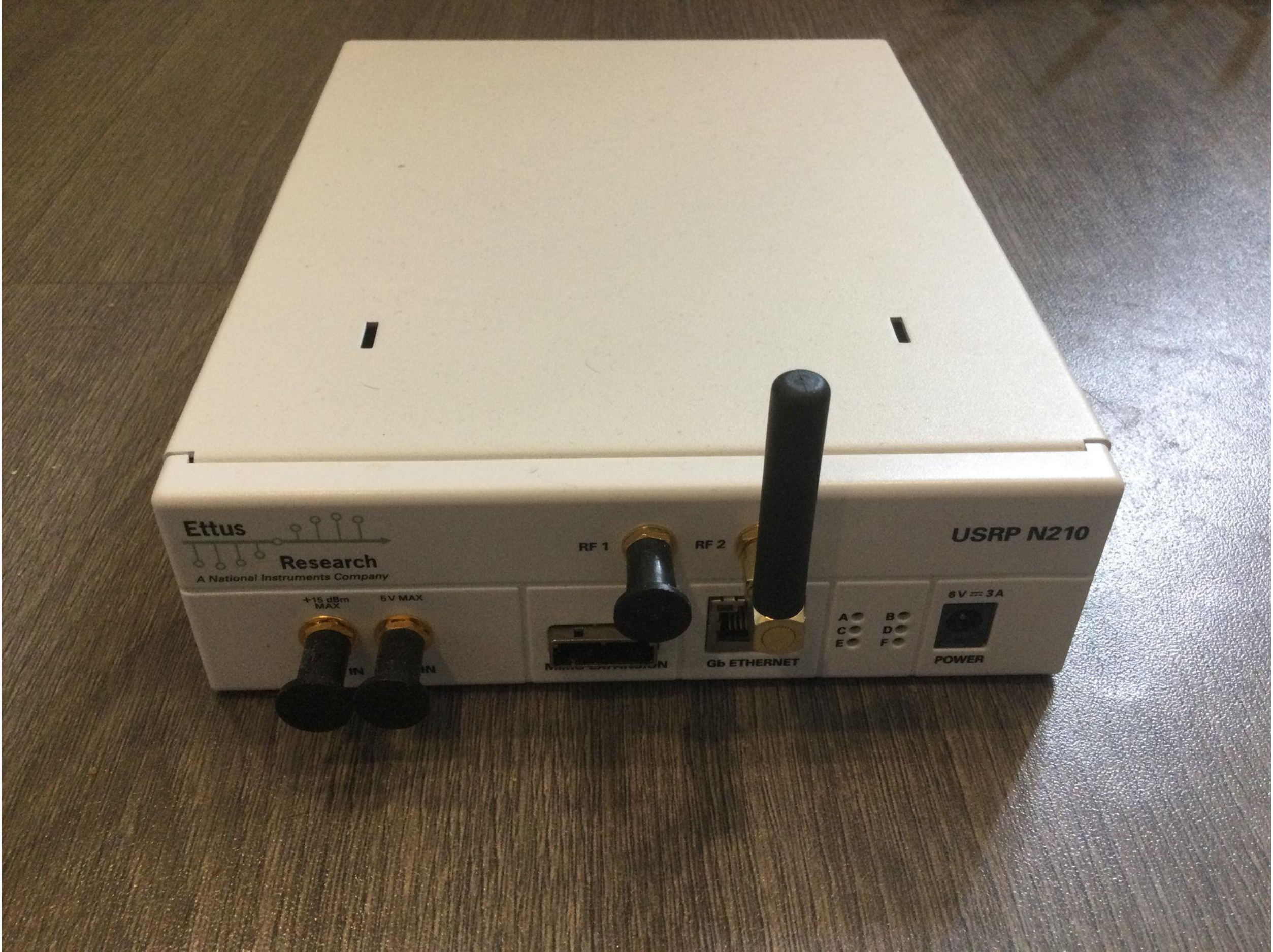}
		\label{}}
	\caption{Experimental devices. (a) LoRa DUTs. (b) USRP N210.}
	\label{fig:exp_dev}
\end{figure}

10 LoRa devices under test (DUTs) are employed. As shown in Fig.~\ref{fig:exp_dev}(a), five of them are LoPy4 and the others are Dragino LoRa shields, and they all use Semtech SX1276 chipsets.
All LoRa DUTs are configured with bandwidth of 125~kHz, carrier frequency of 868.1~MHz and 0.5~second transmission interval. 
As shown in Fig.~\ref{fig:exp_dev}(b), a USRP N210 SDR platform is adopted as the receiver, and its sampling rate is set to 250~kHz. The LoRa DUTs and USRP N210 are placed half a meter apart with no obstacle between them (line-of-sight). Data collection was conducted device by device. Each LoRa DUT has the SF set to 7, 8, and 9 in turn and 3,000 packets are transmitted per SF.

\subsubsection{Dataset Description}
The experimental dataset contains 90,000 packets from 10 DUTs under three SF configurations. 75,000 of them (25,000 for each SF) are used for training, and the rest 15,000 packets (5,000 packets for each SF) for testing. The preamble part of packets with SF 7, 8, and 9 contains IQ samples with lengths of 2,048, 4,096 and 8,192, respectively. The SNR of all the received packets is estimated to be around 70~dB. Therefore, we treat them as noiseless signals and add AWGN to them to simulate different SNR conditions during the test.

\subsubsection{Neural Network Training Details}
The neural networks are trained according to the pipelines described in Section~\ref{sec:training_pipeline}. The SNR uniformly ranges from 0~dB to 40~dB when generating the AWGN channels for both online and offline augmentation. 10\% training packets are randomly selected for validation. Adam is used as the optimizer and the mini-batch size is set to 32~\cite{kingma2014adam}. A learning rate scheduler is adopted. The initial learning rate is 0.001 and is reduced by a factor of 0.2 every time the validation loss stops decreasing for five epochs. The training stops when the validation loss plateaus for ten epochs.
The neural networks are implemented with the Tensorflow and Keras libraries, and trained on a GPU of NVIDIA GeForce GTX 1660.

\subsection{Performance of Different SF Configurations}
\begin{figure}[!t]
	\centering
	\subfloat[]{\includegraphics[width=1.7in]{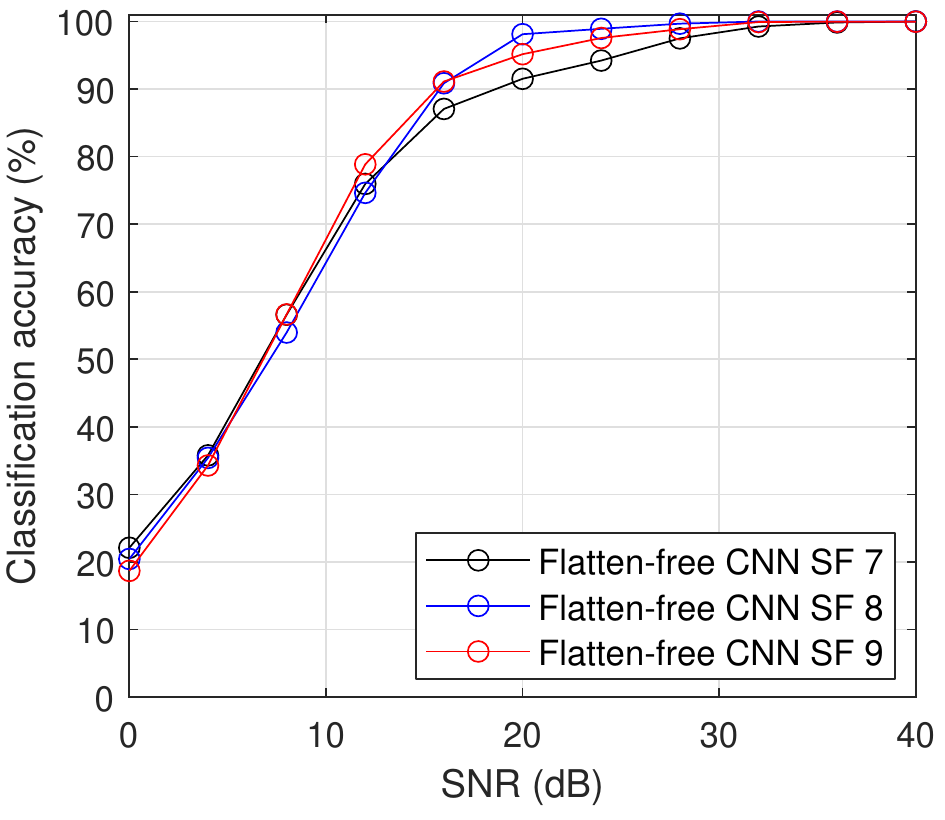}
		\label{}}
	\subfloat[]{\includegraphics[width=1.7in]{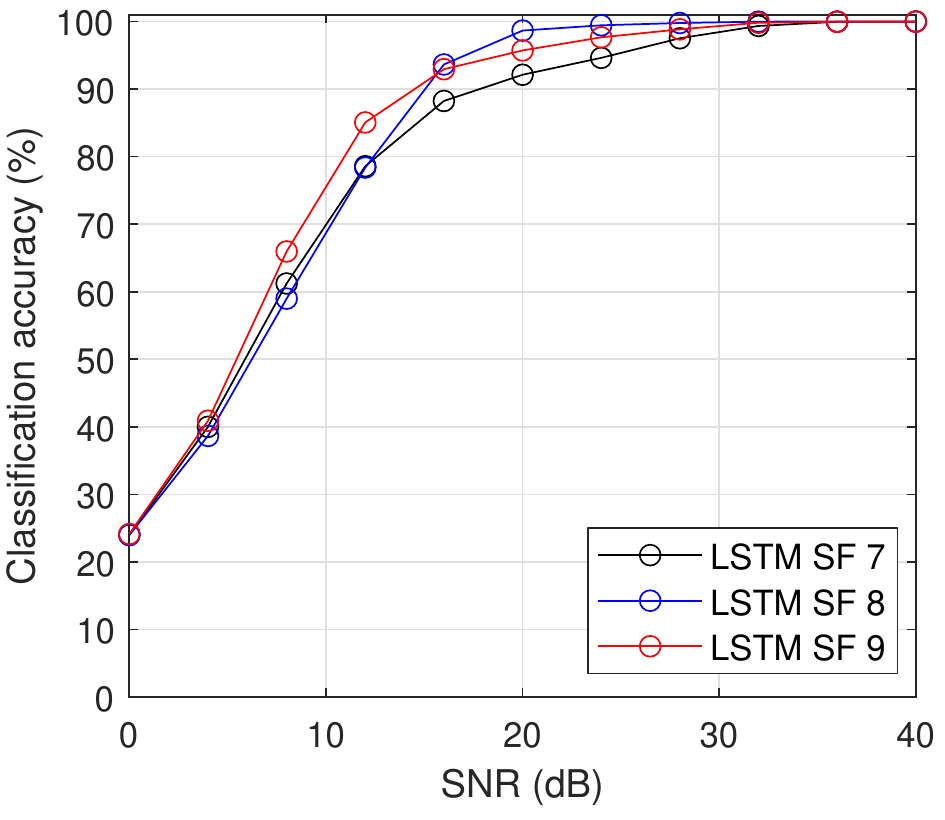}
		\label{}}
		
	\subfloat[]{\includegraphics[width=1.7in]{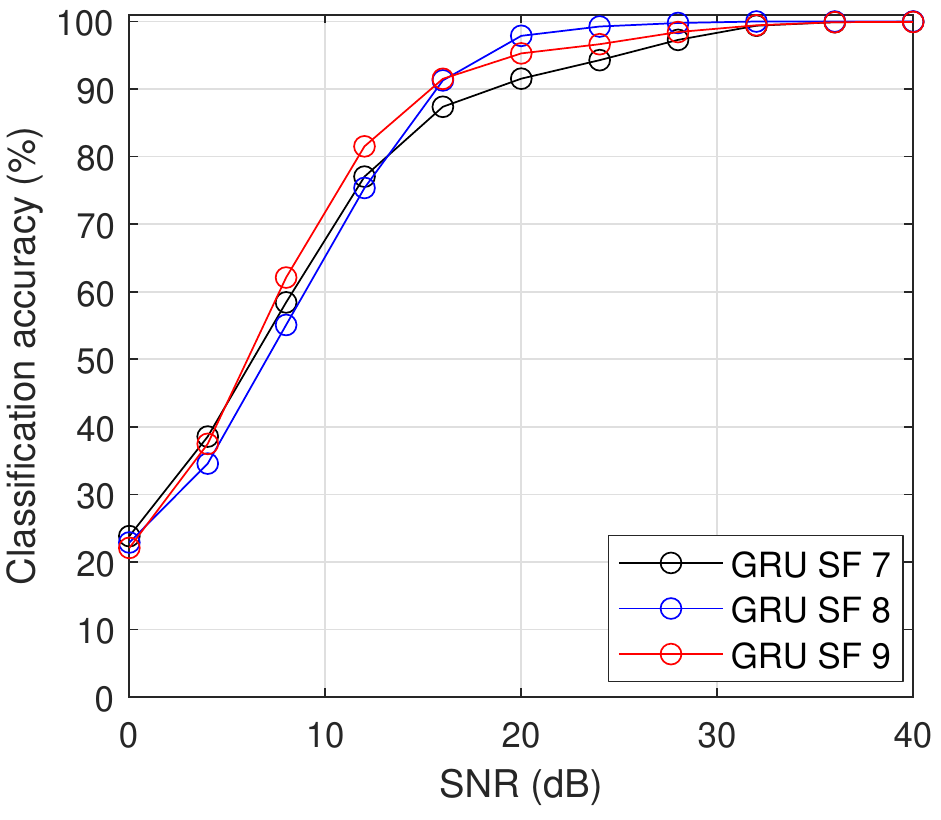}
		\label{}}
	\subfloat[]{\includegraphics[width=1.7in]{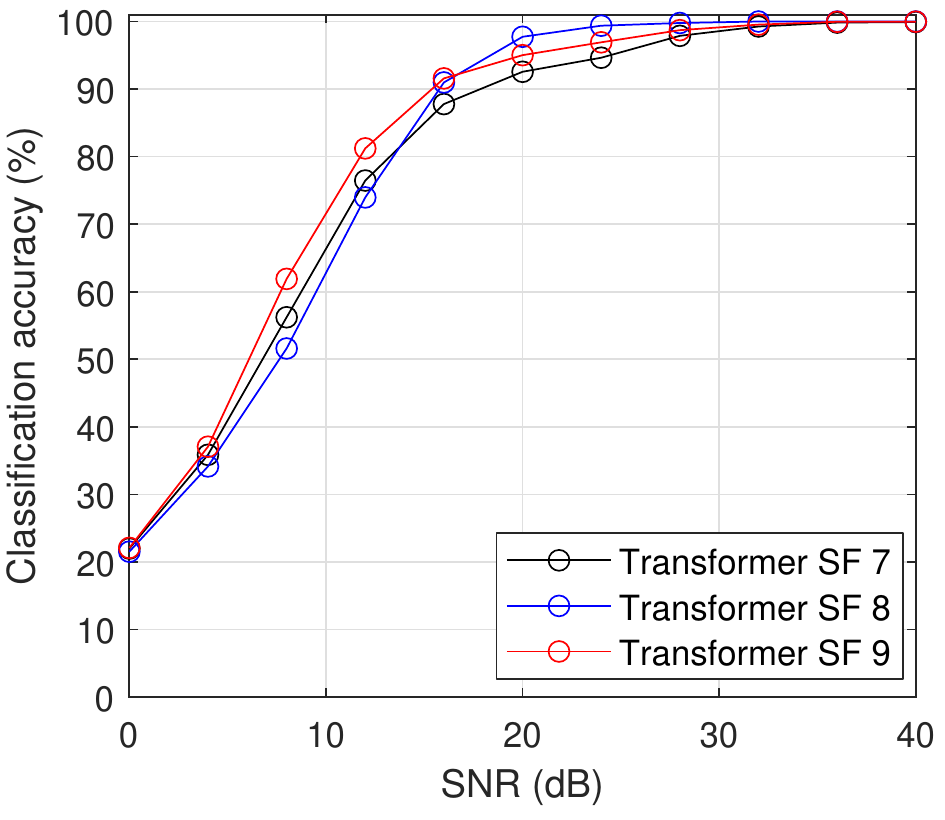}
		\label{}}	
	\caption{Classification accuracy under different SFs. The models are trained with SF 7, 8, 9 signals using online augmentation. Inference with SF 7, 8, 9 signals without the use of multi-packet protocol. (a) Flatten-free CNN. (b) LSTM network. (c) GRU network. (d) Transformer.}
	\label{fig:spreading_factor}
\end{figure}

The proposed flatten-free CNN, LSTM, GRU and transformer models can process channel independent spectrograms under different SF configurations (Fig.~\ref{fig:channel_ind_spectrogram}). Their classification results are shown in Fig.~\ref{fig:spreading_factor}. 
The four sub-figures demonstrate that there is no evident performance gap among the four models. In addition to this, there is also no significant difference when classifying signals of different SFs.

\begin{figure*}[!t]
	\centering
	\subfloat[]{\includegraphics[width=1.4in]{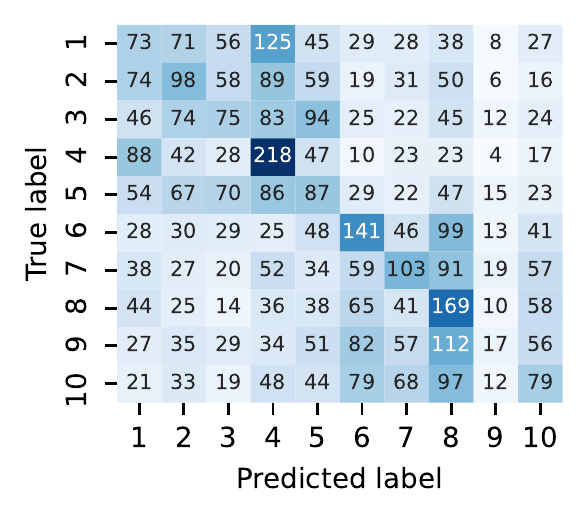}
		\label{}}
	\subfloat[]{\includegraphics[width=1.4in]{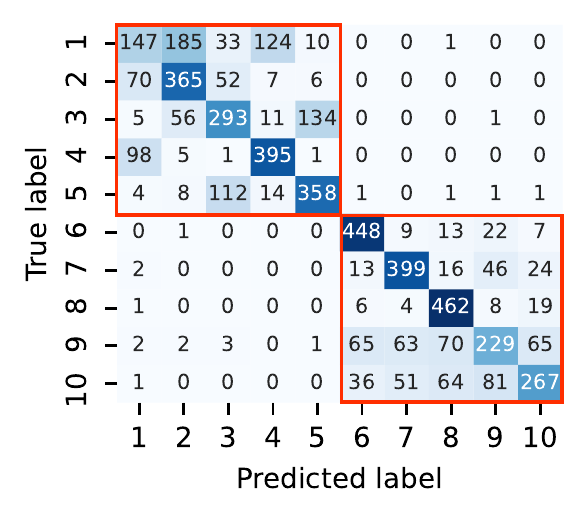}
		\label{}}
	\subfloat[]{\includegraphics[width=1.4in]{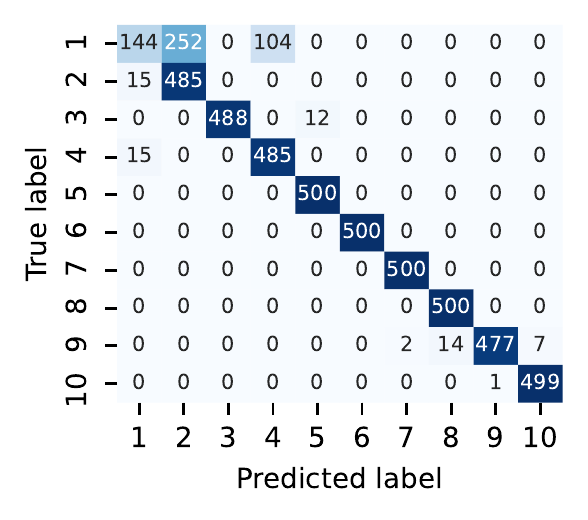}
		\label{}}	
	\subfloat[]{\includegraphics[width=1.4in]{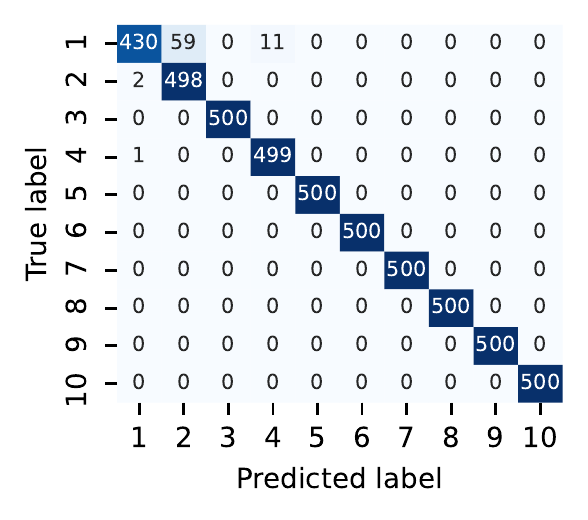}
		\label{}}
	\subfloat[]{\includegraphics[width=1.4in]{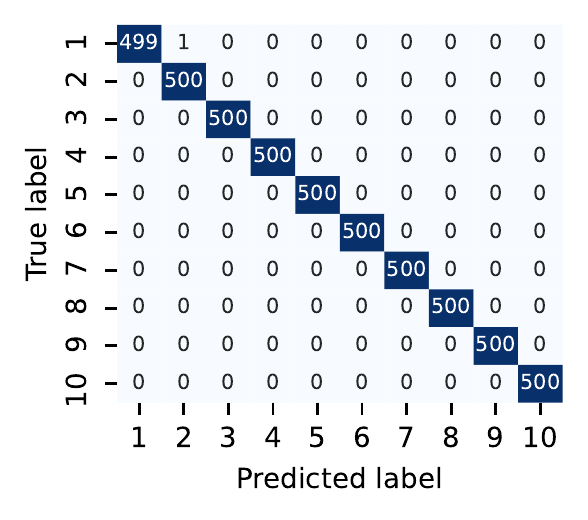}
		\label{}}	
	\caption{Confusion matrices at various SNRs using flatten-free CNN model. The model is trained with SF 7, 8, 9 signals using online augmentation. Inference with SF 7 signals without the use of the multi-packet method. Each device has 500 packets to be classified.  (a) SNR: 0~dB, overall accuracy: 21.20\%. (b) SNR: 10~dB, accuracy: 67.26\%. (c) SNR: 20~dB, accuracy: 91.56\%. (d) SNR: 30~dB, accuracy: 98.52\%. (e) SNR: 40~dB, accuracy: 99.98\%.}
	\label{fig:confusion_matrix}
\end{figure*}

In terms of accuracy, the RFFI performance is nearly perfect when SNR is over 30~dB. 
The accuracy is reduced to 60\% at 10 dB, but the multi-packet inference can increase it to 90\%, which will be shown in Fig.~\ref{fig:multi_pkt_inference}. Fig.~\ref{fig:confusion_matrix} shows the confusion matrices of the flatten-free CNN model at various SNRs, which provides detailed classification results.
Fig.~\ref{fig:confusion_matrix}(a) indicates that the RFFI protocol cannot successfully classify LoRa DUTs at 0~dB. The best case is DUT~4, where 218 out of the 500 packets are correctly classified. The worst case is DUT~9, where only 17 out of 500 packets are correctly classified. When the SNR is increased to 10~dB, as shown in Fig.~\ref{fig:confusion_matrix}(b), we can see that all the predictions are in the red boxes, which means the RFFI protocol is capable of distinguishing the types of DUT. More specifically, a LoPy4 DUT will rarely be classified as a Dragino LoRa shield since their characteristics are different. However, the system still cannot classify LoRa DUTs from the same manufacturer with high accuracy. In contrast, most DUTs can be correctly classified at the higher 20~dB, while only the accuracy of DUT~1 is unsatisfactory. This is probably because DUT~1 has very similar characteristics to DUT~2 and DUT~4. Therefore, misclassification can still occur even at 20~dB. At the nearly ideal 40~dB shown in Fig.~\ref{fig:confusion_matrix}(e), the RFFI protocol achieves perfect classification accuracy for all the DUTs.


\subsection{Comparison of Augmentation Strategies}

We trained each deep learning model (flatten-free CNN, LSTM, GRU and transformer) using online, offline and no augmentation strategies, respectively. For clarity, we only use SF 7 signals during inference to compare the performance of different augmentation strategies, since there is no significant performance gap among different SF settings. 

\begin{figure}[!t]
	\centering
	\subfloat[]{\includegraphics[width=1.7in]{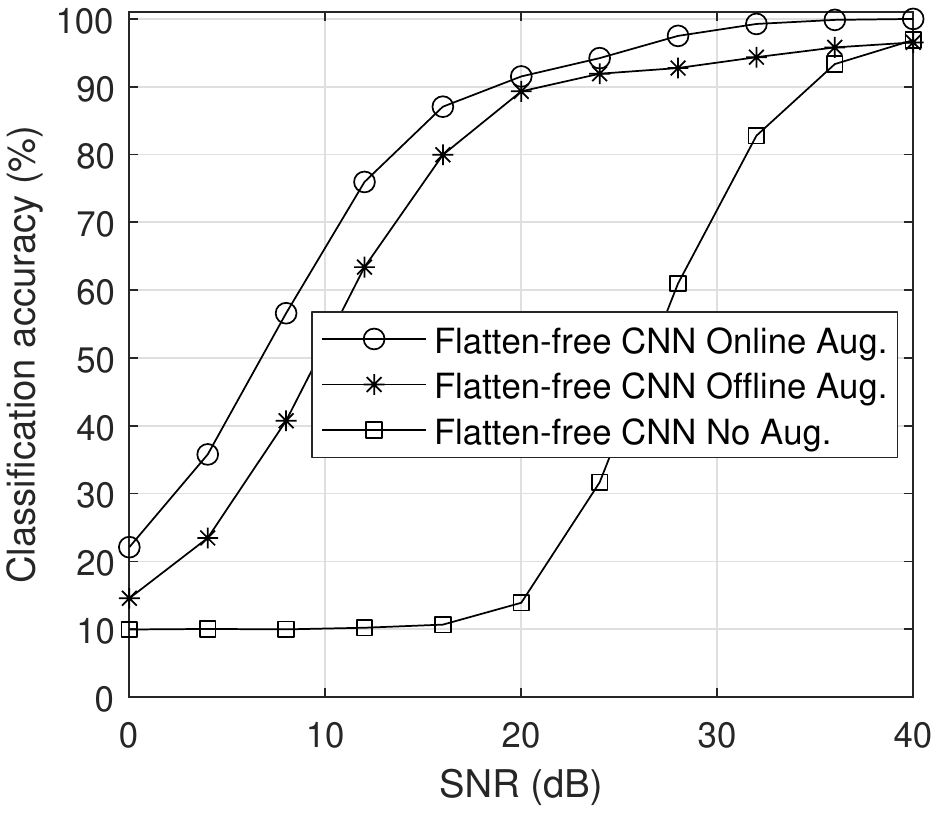}
		\label{}}
	\subfloat[]{\includegraphics[width=1.7in]{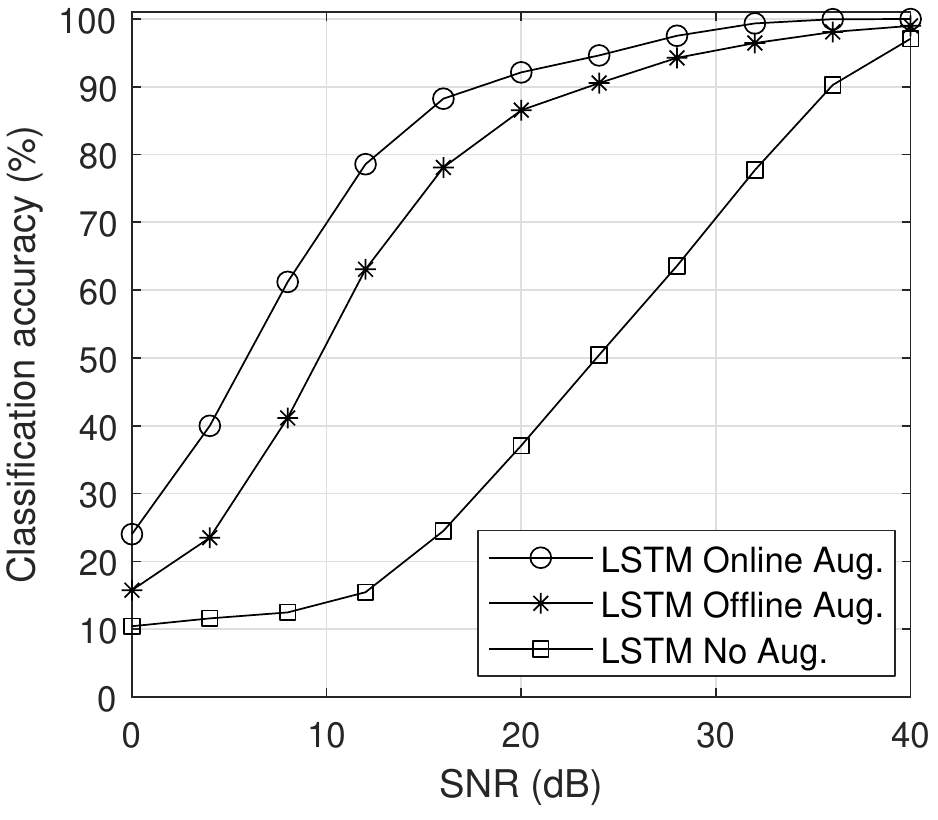}
		\label{}}
		
	\subfloat[]{\includegraphics[width=1.7in]{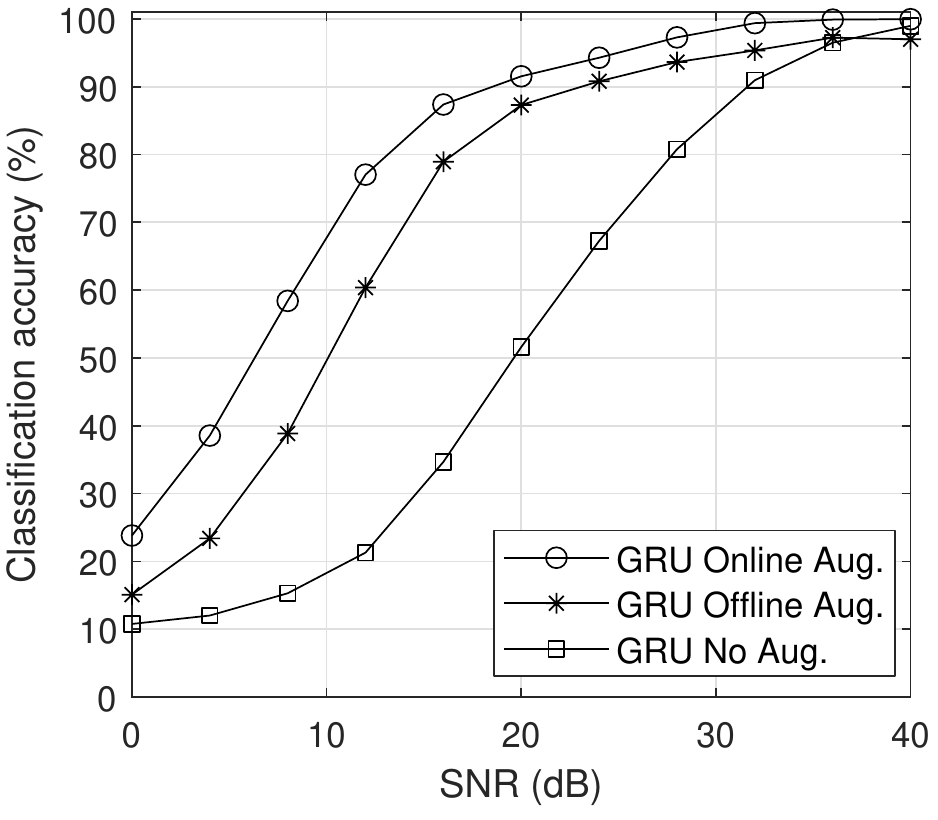}
		\label{}}
	\subfloat[]{\includegraphics[width=1.7in]{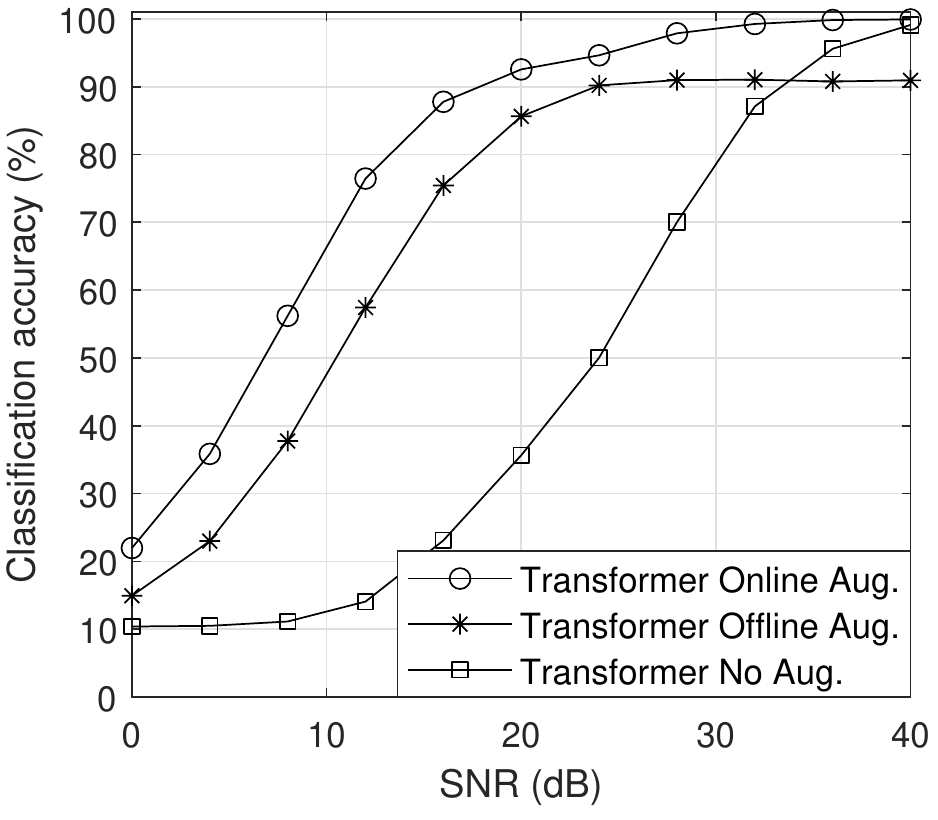}
		\label{}}	
	\caption{Comparison between augmentation strategies. The models are trained with SF 7, 8, 9 signals using online, offline and no augmentation strategies, respectively. Inference with SF 7 signals without the use of the multi-packet inference method. (a) Flatten-free CNN. (b) LSTM network. (c) GRU network. (d) Transformer.}
	\label{fig:augmentation_comparison}
\end{figure}

Their performance at different SNRs is shown in Fig.~\ref{fig:augmentation_comparison}.
The models trained with online augmentation outperform the ones trained with offline augmentation, since the models can learn many more noisy signals during the online augmentation compared to the offline one. We take the training of the flatten-free CNN as an example. The training with online augmentation stops after 140,000 steps. Therefore, the flatten-free CNN has learned $140,000\times32$ (mini-batch size) noisy signals. In contrast, offline augmentation can only provide 75,000 (training set size) noisy signals to the flatten-free CNN for learning. Although replicating the training dataset several times may increase the number of noisy signals to the same level as online augmentation, it will significantly increase the disk storage size and memory requirements. In our case, when using offline training, we need to replicate the dataset around 60 times to match the number of noisy signals provided during online training, which is unaffordable for a large-scale dataset.
It can also be observed that the models trained without any augmentation perform well at high SNRs but poorly at low SNRs. This demonstrates data augmentation must be employed during training, otherwise, the RFFI protocol may not work even at 20~dB. 

Online augmentation requires more training time than the offline one. Take the training of flatten-free CNN as an example, online augmentation costs 200~minutes, while offline costs only 70~minutes. The increasing training time can be tolerated as the neural network can be trained on a powerful workstation or cloud server and then deployed to the gateway.

\subsection{Effect of Multi-packet Inference}
\begin{figure}[!t]
	\centering
	\subfloat[]{\includegraphics[width=1.7in]{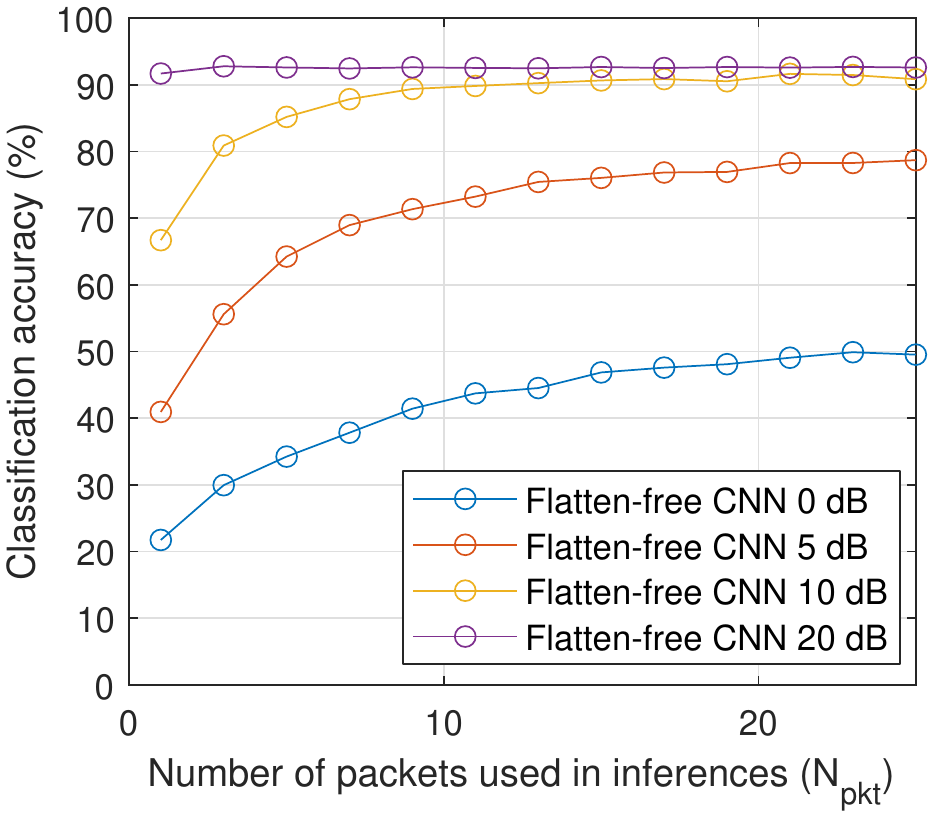}
		\label{}}
	\subfloat[]{\includegraphics[width=1.7in]{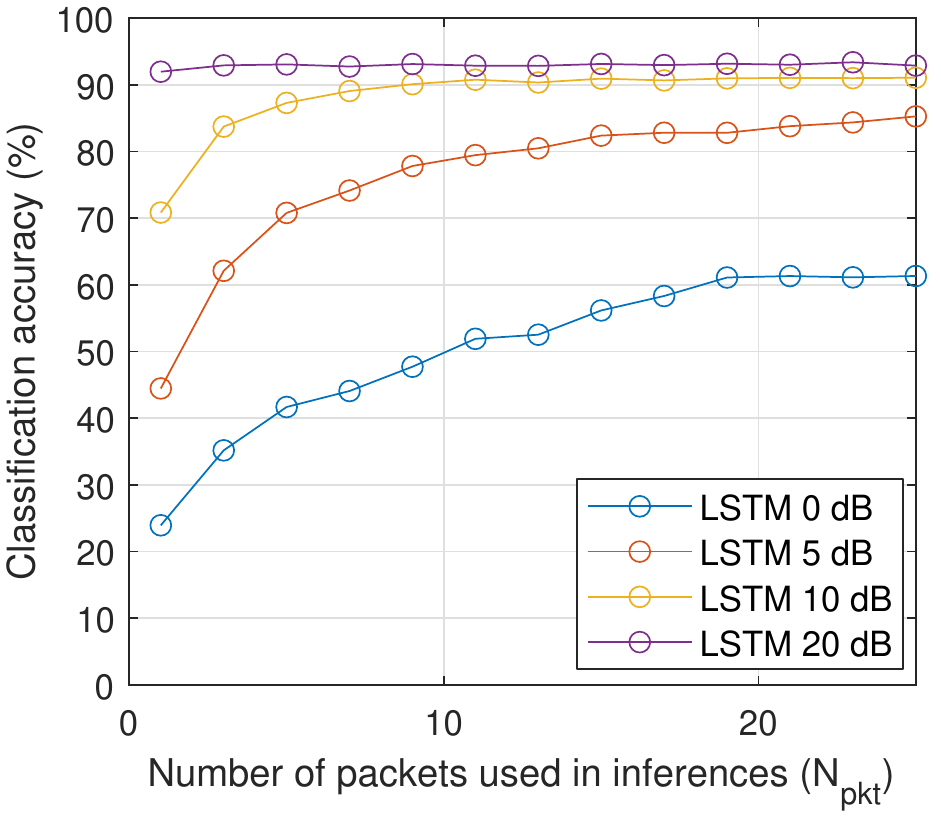}
		\label{}}
		
	\subfloat[]{\includegraphics[width=1.7in]{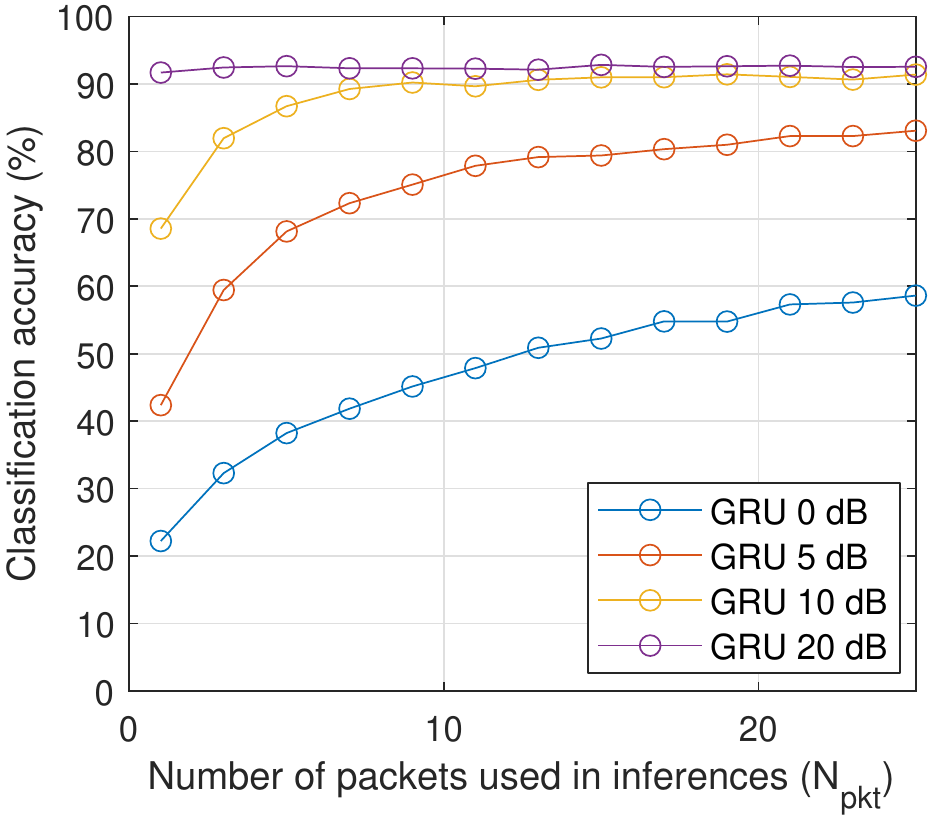}
		\label{}}
	\subfloat[]{\includegraphics[width=1.7in]{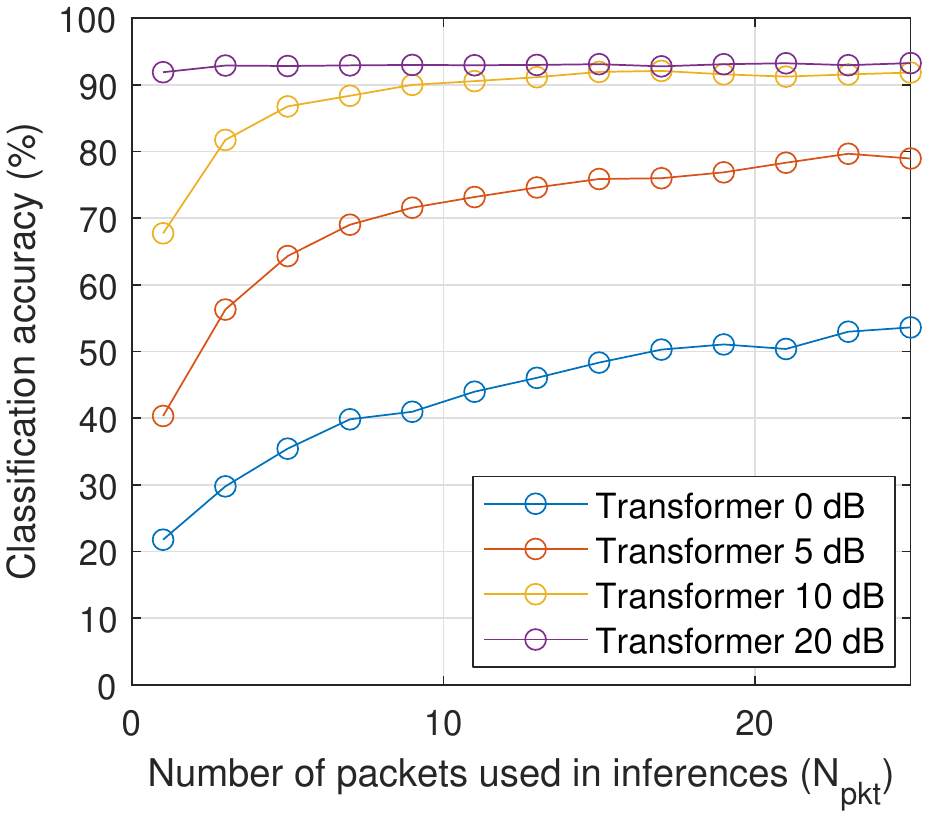}
		\label{}}	
	\caption{Effect of multi-packet inference method. Each model is trained with SF 7, 8, 9 signals using online augmentation. Inference with SF 7 signals. (a) Flatten-free CNN. (b) LSTM network. (c) GRU network. (d) Transformer.}
	\label{fig:multi_pkt_inference}
\end{figure}

The proposed multi-packet inference approach can significantly improve RFFI performance, particularly in low SNR scenarios. The effect of packet number on classification accuracy is studied at four different SNRs, and the results are given in Fig.~\ref{fig:multi_pkt_inference}. 
As shown in the figure, the multi-packet inference can increase the classification accuracy of the four models by over 20\% at 0~dB, 5~dB and 10~dB, while obtaining marginal improvement at 20~dB. This demonstrates multi-packet inference can be adopted in low SNR scenarios to effectively improve system performance.
Furthermore, we can find that the classification accuracy gradually improves as the number of packets increases. However, the improvement is relatively limited after the number of packets exceeds 10. A trade-off should be considered as involving more packets also leads to higher complexity and requires more space to store historical inferences.

\subsection{Complexity Analysis}\label{sec:complexity}


We investigate the model complexity in terms of inference time and number of parameters. Inference time determines the latency of the system, which is an important metric for RFFI. The number of parameters affects the size of the neural network. Larger models require more storage and memory for the embedded platform.
\begin{table}[!t]
  \centering
  \caption{Model complexity information.}
    \begin{tabular}{|l|r|r|}
    \hline
    \multicolumn{1}{|c|}{Neural network} & \multicolumn{1}{c|}{Inference time (SF 7/ 8/ 9)} & \multicolumn{1}{c|}{Parameters} \bigstrut\\
    \hline
    Flatten-free CNN & 28.3/ 32.0/ 40.3 ms  & 675,594  \bigstrut\\
    \hline
    LSTM network & 30.4/ 34.7/ 42.8 ms  & 856,586 \bigstrut\\
    \hline
    GRU network & 27.7/ 31.5 / 36.2 ms  & 644,618  \bigstrut\\
    \hline
    Transformer & 28.7/ 30.3/ 33.8 ms  & 348,938 \bigstrut\\
    \hline\hline
    Slicing CNN & 31.3/ 33.9/ 46.4 ms  & 797,194 \bigstrut\\
    \hline
    \end{tabular}%
  \label{tab:model_complexity}%
\end{table}%
As shown in Table~\ref{tab:model_complexity}, their inference time on SF 7 signals is not much different. The difference is at most 2~ms. However, the gap becomes apparent when classifying the longer SF~9 signals. The transformer costs 33.8~ms, which is 9~ms faster than the slowest LSTM network. Note that the inference time is limited by the deep learning library used and can be further accelerated.
The transformer is also the most lightweight model, with only 348,938 learnable parameters.


The multi-packet inference protocol requires additional storage and computing resources. We need to store $(N_{pkt}-1)$ historical inferences $\mathbf{\hat{p}}$ for each of the $K$ LoRa devices. Therefore, the historical inference database contains $K\times(N_{pkt}-1)$ probability vectors. As each vector contains $K$ probabilities, the system only needs to store $K\times (K \times(N_{pkt}-1))$ floating-point numbers in total. As described in (\ref{equ:multi_pkt}), the multi-packet inference only requires an additional averaging operation, which is low-complexity.

\subsection{Results Summary}

\subsubsection{Comparison Among Neural Networks}
Transformer can achieve competitive performance with the least parameters and the shortest inference time. Fig.~\ref{fig:spreading_factor} shows that the proposed four models are similar in classification accuracy when classifying signals of different SFs. Table~\ref{tab:model_complexity} demonstrates that the transformer is the most lightweight model and its inference speed is also the fastest.
Therefore, transformer is the recommended neural network considering both performance and complexity.

\subsubsection{Solutions to Low SNR RFFI}
The RFFI performance in low SNR scenarios can be improved in two ways. One is applying online augmentation during training to increase model noise robustness, and another is to merge the results of multiple packets/observations during inference. Online augmentation increases the training overhead which can be resolved by training on a cloud server. Using multi-packet inference is effective in low SNR scenarios, which can significantly improve the accuracy by over 20\%.

\section{Comparison with Slicing Technique}\label{sec:slicing}


An alternative solution to the variable-length input problem is the slicing/splitting technique, which is employed in~\cite{merchant2018deep,al2020exposing,soltani2020more,yu2019robust}. In this section, we compare the performance of the slicing/splitting technique with our proposed length-versatile neural networks.

\subsection{Slicing Technique}
Slicing/splitting is to divide the received signal into shorter slices/segments with equal length. During the inference, the CNN infers from each slice individually and the softmax outputs of all the slices are averaged to predict the label. In this scheme, the CNN always accepts equal-length slices as inputs so that the variable-length problem can be avoided. 

\begin{figure}[!t]
	\centering
	\includegraphics[width=3.4in]{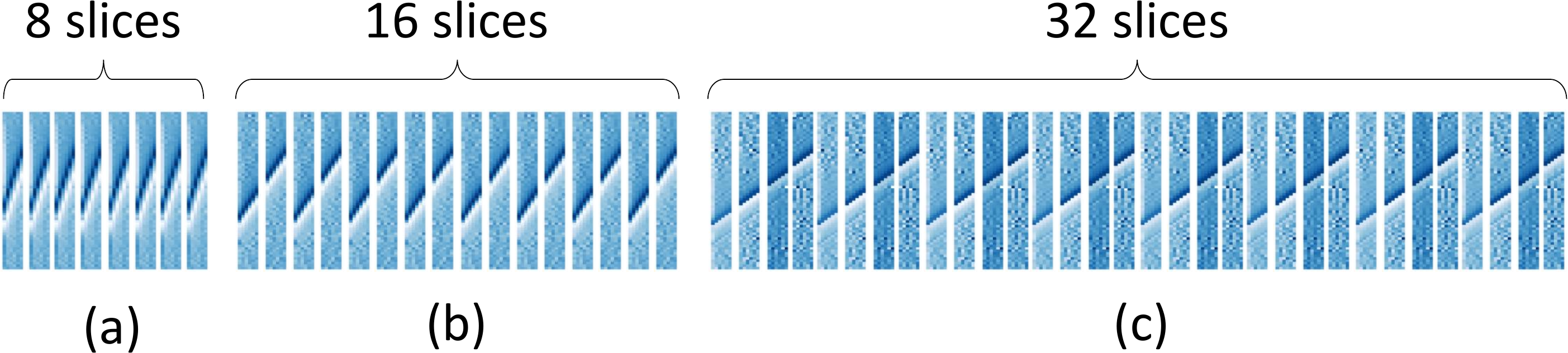}
	\caption{Sliced channel independent spectrograms of LoRa preambles. (a) SF 7 preambles are split into eight slices. (b) SF~8 preambles are split into 16 slices. (c) SF~9 preambles are split into 32 slices.}
	\label{fig:channel_ind_spectrogram_slices}
\end{figure}

In our implementation, we divide the preamble part of each LoRa packet into several slices of 256 IQ samples, and then convert them to $(64, 6)$ channel independent spectrograms as CNN inputs. The SF 7, 8, 9 signals contain 2,048, 4,096 and 8,192 IQ samples, respectively. Therefore, SF 7 preamble is split into eight slices, each corresponding to one LoRa SF 7 preamble. SF 8 and SF 9 preamble parts are divided into 16 and 32 slices, respectively. Each SF 8 slice contains half a LoRa SF 8 preamble and each SF 9 slice contains a quarter of a LoRa SF 9 preamble. The slices can be viewed in Fig~\ref{fig:channel_ind_spectrogram_slices}.

\begin{figure*}[!t]
	\centering
	\includegraphics[width=6.8in]{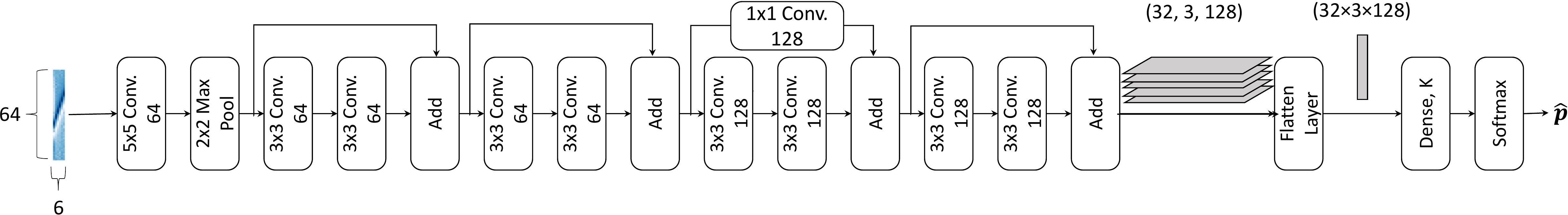}
	\caption{Architecture of the slicing CNN.}
	\label{fig:cnn_slices}
\end{figure*}

We designed an ordinary CNN with a flatten layer to process the slices, this is termed the slicing CNN, whose structure is shown in Fig.~\ref{fig:cnn_slices}.
The slicing CNN has a flatten layer, but in the flatten-free CNN it is replaced with a global average pooling 2D layer, which is the only difference between these two architectures. Therefore, we can assume they have similar learning abilities. The slicing CNN is trained using exactly the same settings as the flatten-free CNN, including optimizer, learning rate, mini-batch size and stop conditions, etc.
Therefore, a fair comparison can be conducted.

\subsection{Performance Comparison}\label{sec:performance_comparison}
\begin{figure}[!t]
	\centering
	\includegraphics[width=3.4in]{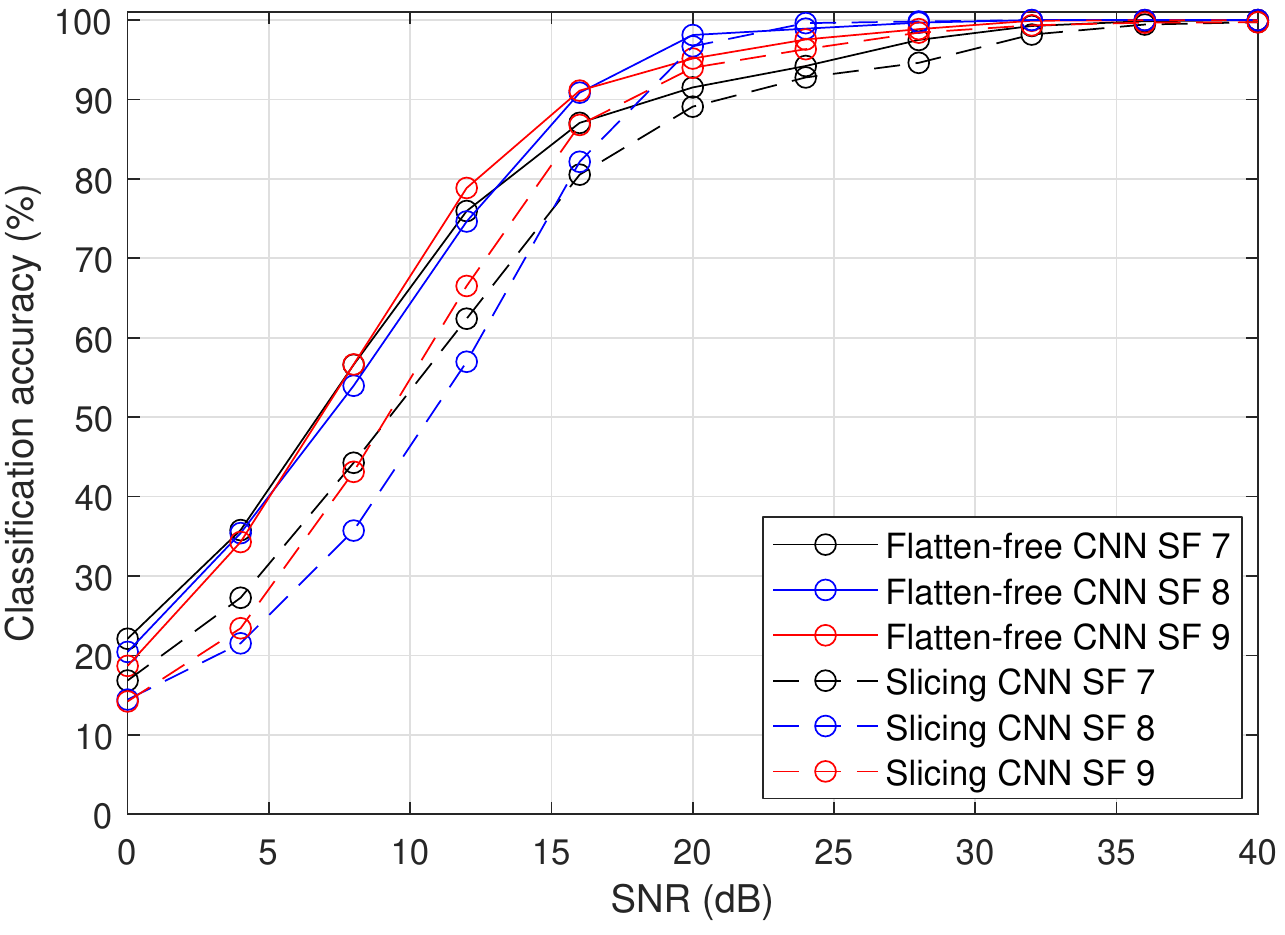}
	\caption{Performance comparison between slicing CNN and flatten-free CNN}
	\label{fig:slicing_comparison}
\end{figure}

The performance comparison between the slicing CNN and the flatten-free CNN is shown in Fig.~\ref{fig:slicing_comparison}.
The flatten-free CNN performs better than the slicing CNN in all the three SF configurations. Although both of them can achieve near-perfect performance when SNRs are above 30~dB, the accuracy of flatten-free CNN is about 10\% higher than the slicing technique at low SNRs. The reason for this performance gap will be discussed in Section~\ref{sec:difference_preambles}. 

We also analyzed the complexity of the slicing CNN, which is included in Table~\ref{tab:model_complexity}. It has 797,194 parameters in total. The inference time (sum of all slices) of SF 7, 8 and 9 signals is 31.3, 33.9 and 46.4~ms, respectively. Compared to the complexity information of flatten-free CNN shown in Table~\ref{tab:model_complexity}, the slicing CNN requires more storage space and longer inference time.


\subsection{Characteristic Difference Among Preambles}\label{sec:difference_preambles}


In this section, we discuss the reason for the performance gap between the slicing and our proposed length-versatile techniques. We experimentally show that the characteristics of the eight LoRa preambles are different from each other, even though their payloads are the same. Therefore, when using length-versatile techniques, the input to the flatten-free CNN contains all eight preambles with different characteristics. Then the flatten-free CNN can learn the correlation among the eight feature-different preambles. In contrast, the slicing technique only processes a short section of the signal each time.

We take SF 7 signals as an example to show that the eight preambles in a LoRa packet are different in characteristics. As shown in Fig.~\ref{fig:channel_ind_spectrogram_slices}, the SF 7 channel independent spectrogram is splitting into eight slices, each slice corresponding to a LoRa preamble. We then trained eight CNNs with preambles in different positions. More specifically, CNN~1 is trained with the first preamble of all the training packets, while CNN~2 is trained with the second preamble, and so forth. Then we use the preamble in different positions to test the eight trained CNN models. The classification results are given in Table~\ref{tab:preamble_difference}.
\begin{table}[!t]
  \centering
  \caption{Characteristic difference among preambles.}
    \scalebox{0.9}{
    \begin{tabular}{|l|r|r|r|r|r|r|r|r|}
    \multicolumn{9}{c}{Test Preamble} \bigstrut[b]\\
    \hline
       & 1st  & 2nd  & 3rd  & 4th  & 5th  & 6th  & 7th  & 8th \bigstrut\\
    \hline
    CNN 1 & \textbf{100\%} & 49\% & 47\% & 41\% & 39\% & 38\% & 40\% & 45\% \bigstrut\\
    \hline
    CNN 2 & 50\% & \textbf{93\%} & \textbf{93\%} & \textbf{93\%} & 79\% & 54\% & 51\% & 54\% \bigstrut\\
    \hline
    CNN 3 & 45\% & 87\% & \textbf{93\%} & \textbf{92\%} & \textbf{90\%} & 82\% & 66\% & 66\% \bigstrut\\
    \hline
    CNN 4 & 50\% & 60\% & 89\% & \textbf{92\%} & \textbf{94\%} & \textbf{93\%} & 77\% & 74\% \bigstrut\\
    \hline
    CNN 5 & 40\% & 58\% & \textbf{92\%} & \textbf{90\%} & \textbf{92\%} & \textbf{93\%} & \textbf{92\%} & 86\% \bigstrut\\
    \hline
    CNN 6 & 54\% & 69\% & 86\% & 76\% & 84\% & \textbf{91\%} & \textbf{92\%} & 93\% \bigstrut\\
    \hline
    CNN 7 & 60\% & 71\% & 70\% & 77\% & 83\% & 85\% & \textbf{92\%} & \textbf{93\%} \bigstrut\\
    \hline
    CNN 8 & 67\% & 77\% & 80\% & 73\% & 73\% & 76\% & 84\% & \textbf{91\%} \bigstrut\\
    \hline
    \end{tabular}%
    }
  \label{tab:preamble_difference}%
\end{table}%

The results demonstrate that the CNN trained with preambles in a specific position cannot accurately classify the preambles in other positions. For instance, CNN~1 model is trained with the first preamble and it only achieves 49\% accuracy in classifying the second preamble, which reveals that the first preamble has different characteristics from the second one. 

\begin{figure}[!t]
	\centering
	\includegraphics[width=3.4in]{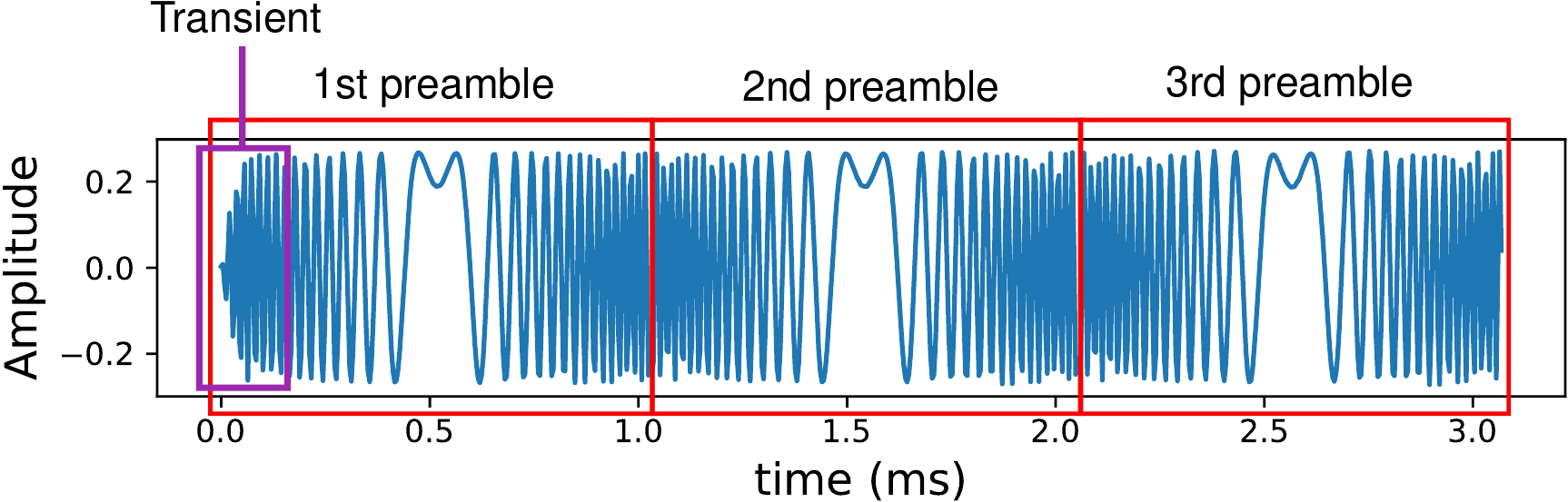}
	\caption{The first three preambles in a LoRa packet. The waveform in the purple box is the signal transient part.}
	\label{fig:preamble_transient}
\end{figure}

Fig.~\ref{fig:preamble_transient} shows the waveform of the first and second LoRa preambles.
They are almost identical except for the part in the purple box. The start of the first preamble shows a gradual increase in signal amplitude. We presume this is the signal transient part that is generated when the hardware components are powered on, which serves as features in traditional RFFI techniques~\cite{danev2009transient,danev2009physical}. Typically, these transient-based methods require equipment with a high sampling rate to collect sufficient IQ samples of the transient part. However, Fig.~\ref{fig:preamble_transient} demonstrates that the relatively low-cost USRP SDR can capture it as well. The first preamble contains signal transient part, which differs from the other preambles. Therefore, as shown in Table~\ref{tab:preamble_difference}, the CNN~1 that is trained with the first preamble does not work well on the others.

Even though the difference from the second to the eighth preamble cannot be observed directly from the waveform, it can be revealed by the classification results in Table~\ref{tab:preamble_difference}. We highlight the accuracy above 90\% and find that they are distributed along the diagonal. This demonstrates that the characteristics of the preamble in different positions are different.

The above discussion shows that the characteristics of slices (preambles) are different even though each preamble is modulated with the same payload.
Therefore, the flatten-free CNN can learn the difference/correlation among them since it processes eight preambles simultaneously. In contrast, the slicing CNN only receives a short slice each time, and all per-slice inferences are merged by simply averaging. This disables the CNN's ability to learn the correlation among feature-different preambles, which results in worse performance of the slicing CNN shown in Fig.~\ref{fig:slicing_comparison}.



\section{Related Work}
RFFI identifies wireless devices by analyzing the characteristics of  received signals. Traditionally, this is achieved by human-designed feature extraction algorithms in combination with conventional machine learning classification models. The manually extracted features can be the estimated hardware characteristics, such as the estimated CFO, IQ imbalance, phase error~\cite{shi2011improved,brik2008wireless,zhuang2018fbsleuth}, power amplifier non-linearity~\cite{polak2011identifying} and beam pattern~\cite{balakrishnan2019physical}. They can also be the statistics of the received signal, such as kurtosis, spectral flatness and brightness~\cite{joo2020hold,satija2018specific}. In addition to these, some domain transformation results and variants of constellation figures can serve as features too. For instance, the Hilbert-Huang spectrum~\cite{zhang2016specific}, power density spectrum~\cite{wang2016physical, danev2009physical} and differential constellation trace figure (DCTF)~\cite{peng2018design} are examples. These manually designed features are extracted from the received signals and then classified using support vector machine (SVM) or k-nearest neighbor (kNN) models.
Recently, deep learning models demonstrated remarkable performance for RFFI~\cite{zhang2021radio,liu2021bidirectional,he2016deep,al2021deeplora,shen2021infocom,shen2021jsac,peng2019deep,merchant2018deep,yu2019robust,xie2021generalizable,roy2019rfal,soltani2020more,al2020exposing,ozturk2020rf,qian2021specific,das2018deep}. They can automatically learn excellent features after sufficient training, thus eliminating the need for feature engineering.

The current deep learning-based RFFI protocols still have some defects, one of which is that many RFFI protocols can only handle fixed-length signals. CNN and DNN are widely used because of their excellent feature extraction capability~\cite{yu2019robust,merchant2018deep,al2021deeplora,al2020exposing,soltani2020more,ozturk2020rf,roy2019rfal,qian2021specific,shen2021jsac,das2018deep}. However, they require input signals of fixed size~\cite{al2020exposing}. Therefore, some work uses fixed-length signals as model inputs~\cite{ozturk2020rf,roy2019rfal,qian2021specific,shen2021jsac,hanna2020open,lee2021deep}, e.g., the preamble. This however limits the suitable data and rules out the payload part which usually has variable lengths.
Other research has proposed the slicing/splitting technique that divides the signal into segments with the same length~\cite{yu2019robust,merchant2018deep,al2020exposing,soltani2020more}. However, the slicing/splitting technique can lead to performance degradation according to the results shown in Section~\ref{sec:slicing}.

RFFI performance in low SNR scenarios is also challenging. One improvement approach is to design noise-resilient signal representations as model inputs.
Ozturk~\etal found that time-frequency data (spectrograms) is more resilient to noise than time-series data~\cite{ozturk2020rf}. Xing~\etal proposed a stacking algorithm for direct sequence spread spectrum (DSSS) systems to improve signal quality~\cite{xing2018radio}.  
Alternatively, we can also obtain noise robustness by enhancing the capability of the deep learning model.
Data augmentation is a popular approach to train a channel-agnostic RFFI model~\cite{al2021deeplora,soltani2020more,merchant2019enhanced}, which can also be leveraged to improve the model noise robustness. However, specific analysis on the effect of data augmentation in low SNR scenarios is still missing.

\section{Conclusion}
The research reported here aims to tackle the variable input size and low SNR problems in deep learning-based RFFI protocols. We use LoRa as a case study because it suffers from both of these challenges.
LoRaWAN ADR mechanism results in the variable lengths of LoRa preambles, which requires the model to be able to process inputs of different lengths. Therefore, we present four length-versatile neural networks, namely flatten-free CNN, LSTM, GRU and transformer. LoRa is an LPWAN technology whose received signals may have low SNR, making LoRa-RFFI challenging. Data augmentation can be an efficient approach to boost model noise robustness. We compare the performance of online, offline and no augmentation strategies and found that online augmentation outperforms the others. Furthermore, we leverage a multi-packet inference approach that can considerably improve the system performance in low SNR scenarios. Experiments involving 10 LoRa devices and a USRP N210 SDR were carried out for evaluation. The results show that online augmentation and multi-packet inference is effective in improving the RFFI performance in low SNR conditions.

\bibliographystyle{IEEEtran}
\bibliography{IEEEabrv,mybibfile}

\begin{thebibliography}{10}
\providecommand{\url}[1]{#1}
\csname url@samestyle\endcsname
\providecommand{\newblock}{\relax}
\providecommand{\bibinfo}[2]{#2}
\providecommand{\BIBentrySTDinterwordspacing}{\spaceskip=0pt\relax}
\providecommand{\BIBentryALTinterwordstretchfactor}{4}
\providecommand{\BIBentryALTinterwordspacing}{\spaceskip=\fontdimen2\font plus
\BIBentryALTinterwordstretchfactor\fontdimen3\font minus
  \fontdimen4\font\relax}
\providecommand{\BIBforeignlanguage}[2]{{%
\expandafter\ifx\csname l@#1\endcsname\relax
\typeout{** WARNING: IEEEtran.bst: No hyphenation pattern has been}%
\typeout{** loaded for the language `#1'. Using the pattern for}%
\typeout{** the default language instead.}%
\else
\language=\csname l@#1\endcsname
\fi
#2}}
\providecommand{\BIBdecl}{\relax}
\BIBdecl

\bibitem{zhang2021radio}
J.~Zhang, R.~Woods, M.~Sandell, M.~Valkama, A.~Marshall, and J.~Cavallaro,
  ``Radio frequency fingerprint identification for narrowband systems,
  modelling and classification,'' \emph{{IEEE} Trans. Inf. Forensics Security},
  vol.~16, pp. 3974--3987, 2021.

\bibitem{liu2021bidirectional}
M.~Liu, X.~Han, N.~Liu, and L.~Peng, ``Bidirectional {IoT} device
  identification based on radio frequency fingerprint reciprocity,'' in
  \emph{Proc. IEEE Int. Conf. Commun. (ICC)}.\hskip 1em plus 0.5em minus
  0.4em\relax IEEE, 2021, pp. 1--6.

\bibitem{he2016deep}
K.~He, X.~Zhang, S.~Ren, and J.~Sun, ``Deep residual learning for image
  recognition,'' in \emph{Proc. IEEE Conf. Comput. Vision Pattern Recognition
  (CVPR)}, 2016, pp. 770--778.

\bibitem{al2021deeplora}
A.~Al-Shawabka, P.~Pietraski, S.~B~Pattar, F.~Restuccia, and T.~Melodia,
  ``{DeepLoRa}: Fingerprinting {LoRa} devices at scale through deep learning
  and data augmentation,'' in \emph{Proc. ACM Int. Symposium Mob. Ad Hoc Netw.
  Comput. (MobiHoc)}, Shanghai, China, Jul. 2021.

\bibitem{shen2021infocom}
G.~Shen, J.~Zhang, A.~Marshall, L.~Peng, and X.~Wang, ``Radio frequency
  fingerprint identification for {LoRa} using spectrogram and {CNN},'' in
  \emph{Proc. IEEE Int. Conf. Comput. Commun. (INFOCOM)}, Virtual Conference,
  May 2021.

\bibitem{shen2021jsac}
------, ``Radio frequency fingerprint identification for {LoRa} using deep
  learning,'' \emph{{IEEE} J. Sel. Areas Commun.}, vol.~39, no.~8, pp.
  2604--2616, 2021.

\bibitem{peng2019deep}
L.~Peng, J.~Zhang, M.~Liu, and A.~Hu, ``Deep learning based {RF} fingerprint
  identification using differential constellation trace figure,'' \emph{{IEEE}
  Trans. Veh. Technol.}, vol.~69, no.~1, pp. 1091--1095, 2019.

\bibitem{merchant2018deep}
K.~Merchant, S.~Revay, G.~Stantchev, and B.~Nousain, ``Deep learning for {RF}
  device fingerprinting in cognitive communication networks,'' \emph{{IEEE} J.
  Sel. Topics Signal Process.}, vol.~12, no.~1, pp. 160--167, 2018.

\bibitem{yu2019robust}
J.~Yu, A.~Hu, G.~Li, and L.~Peng, ``A robust {RF} fingerprinting approach using
  multisampling convolutional neural network,'' \emph{{IEEE} Internet Things
  J.}, vol.~6, no.~4, pp. 6786--6799, 2019.

\bibitem{xie2021generalizable}
R.~Xie, W.~Xu, Y.~Chen, J.~Yu, A.~Hu, D.~W.~K. Ng, and A.~L. Swindlehurst, ``A
  generalizable model-and-data driven approach for open-set {RFF}
  authentication,'' \emph{{IEEE} Trans. Inf. Forensics Security}, vol.~16, pp.
  4435--4450, 2021.

\bibitem{roy2019rfal}
D.~Roy, T.~Mukherjee, M.~Chatterjee, E.~Blasch, and E.~Pasiliao, ``{RFAL}:
  Adversarial learning for {RF} transmitter identification and
  classification,'' \emph{{IEEE} Trans. on Cogn. Commun. Netw.}, vol.~6, no.~2,
  pp. 783--801, 2019.

\bibitem{soltani2020more}
N.~Soltani, K.~Sankhe, J.~Dy, S.~Ioannidis, and K.~Chowdhury, ``More is better:
  Data augmentation for channel-resilient {RF} fingerprinting,'' \emph{{IEEE}
  Commun. Mag.}, vol.~58, no.~10, pp. 66--72, 2020.

\bibitem{al2020exposing}
A.~Al-Shawabka, F.~Restuccia, S.~D’Oro, T.~Jian, B.~C. Rendon, N.~Soltani,
  J.~Dy, K.~Chowdhury, S.~Ioannidis, and T.~Melodia, ``Exposing the
  fingerprint: Dissecting the impact of the wireless channel on radio
  fingerprinting,'' in \emph{Proc. IEEE Int. Conf. Comput. Commun. (INFOCOM)},
  Jul. 2020, pp. 646--655.

\bibitem{ozturk2020rf}
E.~Ozturk, F.~Erden, and I.~Guvenc, ``{RF}-based {low-SNR} classification of
  {UAVs} using convolutional neural networks,'' \emph{arXiv preprint
  arXiv:2009.05519}, 2020.

\bibitem{qian2021specific}
Y.~Qian, J.~Qi, X.~Kuai, G.~Han, H.~Sun, and S.~Hong, ``Specific emitter
  identification based on multi-level sparse representation in automatic
  identification system,'' \emph{{IEEE} Trans. Inf. Forensics Security},
  vol.~16, pp. 2872--2884, 2021.

\bibitem{das2018deep}
R.~Das, A.~Gadre, S.~Zhang, S.~Kumar, and J.~M. Moura, ``A deep learning
  approach to {IoT} authentication,'' in \emph{Proc. IEEE Int. Conf. Commun.
  (ICC)}, 2018, pp. 1--6.

\bibitem{robyns2017physical}
P.~Robyns, E.~Marin, W.~Lamotte, P.~Quax, D.~Singel{\'e}e, and B.~Preneel,
  ``Physical-layer fingerprinting of {LoRa} devices using supervised and
  zero-shot learning,'' in \emph{Proc. ACM Conf. Security Privacy Wireless
  Mobile Netw. (WiSec)}, 2017, pp. 58--63.

\bibitem{hanna2020open}
S.~Hanna, S.~Karunaratne, and D.~Cabric, ``Open set wireless transmitter
  authorization: Deep learning approaches and dataset considerations,''
  \emph{{IEEE} Trans. on Cogn. Commun. Netw.}, vol.~7, no.~1, pp. 59--72, 2020.

\bibitem{merchant2019enhanced}
K.~Merchant and B.~Nousain, ``Enhanced {RF} fingerprinting for {IoT} devices
  with recurrent neural networks,'' in \emph{Proc. IEEE Mil. Commun. Conf.
  (MILCOM)}, 2019, pp. 590--597.

\bibitem{shen2021asilomar}
\BIBentryALTinterwordspacing
G.~Shen, J.~Zhang, A.~Marshall, M.~Valkama, and J.~Cavallaro, ``Radio frequency
  fingerprint identification for security in low-cost {IoT} devices,'' in
  \emph{Proc. Asilomar Conference on Signals, Systems, and Computers}, 2021.
  [Online]. Available: \url{https://arxiv.org/abs/2111.14275}
\BIBentrySTDinterwordspacing

\bibitem{loraADR}
``Understanding {ADR},''
  \url{https://lora-developers.semtech.com/uploads/documents/files/Understanding_LoRa_Adaptive_Data_Rate_Downloadable.pdf}
  Accessed Oct. 7, 2021.

\bibitem{shen2021towards}
G.~Shen, J.~Zhang, A.~Marshall, and J.~Cavallaro, ``Towards scalable and
  channel-robust radio frequency fingerprint identification for {LoRa},''
  \emph{arXiv preprint}, 2021.

\bibitem{hochreiter1997long}
S.~Hochreiter and J.~Schmidhuber, ``Long short-term memory,'' \emph{Neural
  computation}, vol.~9, no.~8, pp. 1735--1780, 1997.

\bibitem{chung2014empirical}
J.~Chung, C.~Gulcehre, K.~Cho, and Y.~Bengio, ``Empirical evaluation of gated
  recurrent neural networks on sequence modeling,'' \emph{arXiv preprint
  arXiv:1412.3555}, 2014.

\bibitem{vaswani2017attention}
A.~Vaswani, N.~Shazeer, N.~Parmar, J.~Uszkoreit, L.~Jones, A.~N. Gomez,
  {\L}.~Kaiser, and I.~Polosukhin, ``Attention is all you need,'' in
  \emph{Advances in neural information processing systems}, 2017, pp.
  5998--6008.

\bibitem{kingma2014adam}
D.~P. Kingma and J.~Ba, ``Adam: A method for stochastic optimization,''
  \emph{arXiv preprint arXiv:1412.6980}, 2014.

\bibitem{danev2009transient}
B.~Danev and S.~Capkun, ``Transient-based identification of wireless sensor
  nodes,'' in \emph{Proc. ACM/IEEE Int. Conf. Inf. Process. Sensor Netw.
  (IPSN)}, NW Washington, DC, USA, 2009, pp. 25--36.

\bibitem{danev2009physical}
B.~Danev, T.~S. Heydt-Benjamin, and S.~Capkun, ``Physical-layer identification
  of {RFID} devices.'' in \emph{Proc. USENIX Security Symposium}, 2009, pp.
  199--214.

\bibitem{shi2011improved}
Y.~Shi and M.~A. Jensen, ``Improved radiometric identification of wireless
  devices using {MIMO} transmission,'' \emph{{IEEE} Trans. Inf. Forensics
  Security}, vol.~6, no.~4, pp. 1346--1354, 2011.

\bibitem{brik2008wireless}
V.~Brik, S.~Banerjee, M.~Gruteser, and S.~Oh, ``Wireless device identification
  with radiometric signatures,'' in \emph{Proc. Int. Conf. Mobile Comput. Netw.
  (MobiCom)}, San Francisco, CA, USA, Sep. 2008, pp. 116--127.

\bibitem{zhuang2018fbsleuth}
Z.~Zhuang, X.~Ji, T.~Zhang, J.~Zhang, W.~Xu, Z.~Li, and Y.~Liu, ``{Fbsleuth}:
  Fake base station forensics via radio frequency fingerprinting,'' in
  \emph{Proc. 2018 ACM Asia Conf. Comput. Commun. Secur.}, 2018, pp. 261--272.

\bibitem{polak2011identifying}
A.~C. Polak, S.~Dolatshahi, and D.~L. Goeckel, ``Identifying wireless users via
  transmitter imperfections,'' \emph{{IEEE} J. Sel. Areas Commun.}, vol.~29,
  no.~7, pp. 1469--1479, 2011.

\bibitem{balakrishnan2019physical}
S.~Balakrishnan, S.~Gupta, A.~Bhuyan, P.~Wang, D.~Koutsonikolas, and Z.~Sun,
  ``Physical layer identification based on spatial--temporal beam features for
  millimeter-wave wireless networks,'' \emph{{IEEE} Trans. Inf. Forensics
  Security}, vol.~15, pp. 1831--1845, 2019.

\bibitem{joo2020hold}
K.~Joo, W.~Choi, and D.~H. Lee, ``Hold the door! fingerprinting your car key to
  prevent keyless entry car theft,'' in \emph{Proc. Netw. Distrib. Syst.
  Security Symposium (NDSS)}, Virtual Conference, Feb. 2020.

\bibitem{satija2018specific}
U.~Satija, N.~Trivedi, G.~Biswal, and B.~Ramkumar, ``Specific emitter
  identification based on variational mode decomposition and spectral features
  in single hop and relaying scenarios,'' \emph{{IEEE} Trans. Inf. Forensics
  Security}, vol.~14, no.~3, pp. 581--591, 2018.

\bibitem{zhang2016specific}
J.~Zhang, F.~Wang, O.~A. Dobre, and Z.~Zhong, ``Specific emitter identification
  via {Hilbert--Huang} transform in single-hop and relaying scenarios,''
  \emph{{IEEE} Trans. Inf. Forensics Security}, vol.~11, no.~6, pp. 1192--1205,
  2016.

\bibitem{wang2016physical}
X.~Wang, P.~Hao, and L.~Hanzo, ``Physical-layer authentication for wireless
  security enhancement: Current challenges and future developments,''
  \emph{{IEEE} Commun. Mag.}, vol.~54, no.~6, pp. 152--158, 2016.

\bibitem{peng2018design}
L.~Peng, A.~Hu, J.~Zhang, Y.~Jiang, J.~Yu, and Y.~Yan, ``Design of a hybrid
  {RF} fingerprint extraction and device classification scheme,'' \emph{{IEEE}
  Internet Things J.}, vol.~6, no.~1, pp. 349--360, 2018.

\bibitem{lee2021deep}
W.~Lee, S.~Y. Baek, and S.~H. Kim, ``Deep-learning-aided {RF} fingerprinting
  for {NFC} security,'' \emph{{IEEE} Commun. Mag.}, vol.~59, no.~5, pp.
  96--101, 2021.

\bibitem{xing2018radio}
Y.~Xing, A.~Hu, J.~Zhang, L.~Peng, and G.~Li, ``On radio frequency fingerprint
  identification for {DSSS} systems in low {SNR} scenarios,'' \emph{{IEEE}
  Commun. Lett.}, vol.~22, no.~11, pp. 2326--2329, 2018.

\end{thebibliography}

\end{document}